\pdfoutput=1
\newcommand{\dd}{\mathrm{d}}

\documentclass[aps,reprint,groupedaddress]{revtex4-1}
\usepackage{amsmath}
\usepackage{amsfonts}
\usepackage{graphicx}
\usepackage{epstopdf}
\usepackage{caption}
\usepackage{subcaption}
\usepackage{placeins}
\usepackage{hyperref}

\begin{document}

\title{Standard Model vacuum decay in a de Sitter Background}
\author{Arttu Rajantie}
\affiliation{Department of Physics, Imperial College London, London SW7 2AZ, United Kingdom}
\author{Stephen Stopyra}
\affiliation{Department of Physics, Imperial College London, London SW7 2AZ, United Kingdom}
\date{July 28, 2017}

\begin{abstract}
We present a calculation of thick-wall Coleman-de-Luccia (CdL) bounces in the Standard Model effective potential in a de Sitter background. The calculation is performed including the effect of the bounce back-reaction on the metric, which we compare with the case of a fixed de-Sitter background, and with similar full-backreaction calculation in a model polynomial potential. The results show that the Standard Model potential exhibits non-trivial behavior: rather than a single CdL solution, there are multiple (non-oscillating) bounce solutions which may contribute to the decay rate. All the extra solutions found have higher actions than the largest amplitude solution, and thus would not contribute significantly to the decay rate, but their existence demonstrates that CdL solutions in the Standard Model potential are not unique, and the existence of additional, lower action, solutions cannot be ruled out. This suggests that a better understanding of the appearance and disappearance of CdL solutions in de Sitter space is needed to fully understand the vacuum instability issue in the Standard Model.
\end{abstract}
\maketitle
\section{Introduction\label{sec:intro}}

One of the questions raised by the discovery of the Higgs boson\cite{Aad20121,Chatrchyan201230} has been the implications it has for the stability of the electroweak vacuum. The possibility that the electroweak vacuum might be metastable and vulnerable to spontaneous nucleation of true vacuum bubbles via quantum tunneling was considered even before the discovery of the Higgs boson\cite{SHER1989273}, but the measurements of a Higgs mass around $M_{h} = 125.09 \pm 0.25\rm{ GeV}$ and top quark mass of $173.21 \rm{ GeV}$\cite{Agashe:2014kda} suggest that this may in fact be the real situation in the Standard Model. Of particular note is that these measurements place the Higgs boson in a narrow region of parameter space for which the electroweak vacuum is neither completely stable, nor so unstable that it should have already decayed in the lifetime of the universe\cite{Buttazzo2013}. That the Higgs and top quark masses lie in this narrow metastability region may indicate new physics which stabilizes the potential. Consistency with the present day observations of the electroweak vacuum require that no true vacuum bubble is likely to have nucleated in our past light cone.
\\
Vacuum instability in a Minkowski background has been investigated extensively: see references \cite{Alekhin2012214,Degrassi2012,Buttazzo2013} for example, and the effects of gravitational backreaction of the vacuum bubbles has also been studied by many authors\cite{PhysRevD.77.025034,PhysRevD.95.025008,Salvio2016}. The second point - nucleation of true vacuum bubbles during inflation, is a somewhat more difficult question to answer. This question has been addressed by many authors\cite{Kobakhidze2013130,Branchina2014,PhysRevLett.112.201801,PhysRevLett.113.211102,PhysRevLett.115.241301,Calmet:2017hja,Markkanen:2017dlc,Espinosa2015}. The nucleation rate can be computed by a semi-classical evaluation of the path integral for the vacuum-to-vacuum amplitude, which after a Wick rotation to Euclidean space gives an estimate of the vacuum energy, of which the imaginary part yields the rate at which tunneling out of the false vacuum occurs\cite{PhysRevD.21.3305}. This path integral is dominated by its stationary points: in particular the saddle-points which give the dominant contribution to the imaginary part due possessing an unstable direction. The Euclidean action of these bounce solutions determines the leading order contribution to the decay rate, with the solutions with the \emph{smallest} action dominating.
\\
In this paper we will present numerical calculations of the `bounce solutions', that dominate this path integral, in a de Sitter background and using the Standard Model effective potential. The calculation of bounces in a de Sitter background if often simplified by assuming that the background is a fixed de Sitter metric, unaffected by back-reaction of the bounce solution. This is valid if the difference in energy between the false vacuum, the top of the barrier, and the true vacuum is small compared to the background energy density in the false vacuum, $V_0$. However, the depth of the Standard Model effective potential is such that this assumption does not hold for the Hubble rates usually involved in inflationary cosmology, and the flat space bounce solution is known to probe depths only an order of magnitude below the Planck scale\cite{PhysRevD.77.025034,PhysRevD.95.025008,Salvio2016}. Thus, it is possible that backreaction of the bounce solution could have a significant effect on the nucleation rate. For this reason, we compute the bounce solutions including all these backreaction effects, without assuming that the metric is a fixed de Sitter background. We compare these results with the fixed background case.
\\
Additionally, a recent paper by Joti et al. \cite{Joti2017} also considered vacuum instability in a de Sitter background using a perturbative approach. We will compare our results to \cite{Joti2017} in more detail in section \ref{sec:discussion}.
\\
\subsection{Thermal vs Quantum Tunneling effects}
Another useful technique for computing the decay rate of a false vacuum in de Sitter space is to use the Stochastic approach to inflation\cite{Starobinsky1986}, which considers the long wavelength (superhorizon) behavior of the Higgs field as being effectively a classical field which receives stochastic `kicks' from the sub-horizon field modes. This yields a Langevin and associated Fokker-Planck equation, which can be used to compute the probability that a given portion of false vacuum will remain in the false vacuum for some number of e-folds of inflation. Examples of this approach being applied to Higgs stability during inflation can be found in references \cite{PhysRevLett.112.201801,Espinosa2015}.
\\
A question may be asked, however, about the relationship between the Fokker-Planck approach to computing the false vacuum survival probability and the Coleman de Luccia prescription for computing the nucleation rate of true vacuum bubbles. This process is decidedly sub-horizon, and by averaging over sub-horizon modes the Stochastic approach obscures this information. As a specific example of how things become unclear, consider the late time static solution for the probability distribution arising from the Fokker-Planck equation\cite{Starobinsky1986}:
\begin{equation}
p(\phi) = N\exp\left(-\frac{8\pi^2\Delta V(\phi)}{3H_0^4}\right),\label{eq:FPdist}
\end{equation}
where $H_0 = \sqrt{\frac{V_0}{3M_{\rm{P}}^2}}$ is the Hubble rate, $\Delta V(\phi) = V(\phi) - V_0$ the difference between $V(\phi)$ and the false vacuum potential energy, $V_0 = V(\phi_{\rm{fv}})$.\\

The result of this stochastic analysis can be compared to the Coleman de Luccia (CdL) prescription (see \cite{PhysRevD.21.3305}) for computing the rate of true vacuum bubble nucleation, which in principle should include all the sub-horizon effects. The CdL prescription says that, analogous to flat space, the rate of bubble nucleation in a false vacuum is determined by the action of so called `bounce' solutions to the Euclideanized equation of motion (this will be discussed in more detail in section \ref{sec:bounces}).\\

There is a trivial solution of the Euclidean equations of motion consisting of the field sitting at the top of the barrier - this is known as the Hawking-Moss instanton\cite{HAWKING198235}. Its action, when $|\Delta V(\phi)|\ll V_0$ predicts a decay $\Gamma \propto e^{-B}$ with $B$ matching the exponent in Eq. (\ref{eq:FPdist}). However, there are other, non-trivial solutions - known as Coleman de Luccia (CdL) bounces. These are analogous to the bounce solution that describes tunneling in a Minkowski background\cite{PhysRevD.15.2929,PhysRevD.16.1762}, however, there are important differences - the bounce solutions in flat space must approach the false vacuum at infinity, while the bounce solutions in de-Sitter space exist on a compact Euclidean manifold and do not reach the false vacuum (as we will see in section \ref{sec:bounces}).
The interpretation of these bounce solutions was discussed by Brown and Weinberg\cite{PhysRevD.76.064003}: the tunneling procedure can be regarded as a usual quantum tunneling procedure taking place in a thermal bath of Hawking-Radiation at the Gibbons-Hawking temperature of de-Sitter space\cite{PhysRevD.15.2752}. This explains the exponent in Eq. (\ref{eq:FPdist}): it represents the probability that thermal fluctuations will lift an entire Hubble-volume sphere (volume $\frac{4\pi}{3}\left(\frac{1}{H_0}\right)^3$) from the false vacuum to the top of the barrier of the potential, at the Gibbons-Hawking temperature, $T_{\rm{GH}} = \frac{H_0}{2\pi}$ of the Horizon. The CdL solutions, on the other hand, can be interpreted as tunneling proceeding by thermal excitation pushing the Higgs field partially up the barrier, and then the field tunneling the rest of the way through. Crucially, the Hawking-Moss and CdL solutions describe the \emph{average} effect of many possible routes through the barrier, either through thermal fluctuation, quantum tunneling, or a combination of the two. This is why they appear to describe homogeneous excitations of an entire Hubble volume.\\
\subsection{Existence and uniqueness of bounce solutions}
There is, however, an issue with the Hawking-Moss solution, and thus presumably the Fokker-Planck analysis that appears to reproduce it. Coleman showed that bounce solutions must have one and only one negative eigenvalue in the spectrum of linear field fluctuations about them\cite{COLEMAN1988178} (strictly speaking, as Coleman stated, this analysis does not apply to de Sitter bounces, however, Brown and Weinberg\cite{PhysRevD.76.064003} argued that the conclusion is the same). The Hawking-Moss solution, however, acquires additional negative modes below a certain Hubble rate\cite{PhysRevD.76.064003,PhysRevD.63.123514}, $H_{0\rm{crit}}$, implying that for low Hubble rates, it may not describe the tunneling process. If this is the case, then it would be expected that a CdL solution should control tunneling instead. Indeed, for $H_0 < H_{0\rm{crit}}$, it can be shown that a CdL bounce always exists (provided the potential has a barrier)\cite{PhysRevD.69.063518}. Above this threshold, the existence of CdL bounces is not guaranteed. If there are no CdL bounces for $H_0 > H_{0\rm{crit}}$, then the Hawking-Moss solution controls vacuum decay, and it would be expected by continuity that the CdL bounce merges smoothly with the Hawking-Moss solution as $H_0$ crosses $H_{0\rm{crit}}$ from below. We will argue in this paper, however, that if CdL solutions do exist above this threshold, there \emph{must} be more than one of them.
\\
This brings us to the subject of this paper - CdL solutions \emph{do} exist for $H_0 > H_{0\rm{crit}}$ in the Standard Model effective potential, and thus there is not one but \emph{multiple} CdL solutions contributing to vacuum instability in the Standard Model. We find examples of these extra solutions and compute their Euclidean actions to assess their relevance to tunneling in the Standard Model.
\\
\subsection{Overview}
We will address these questions by computing (numerically) bounce solutions in the Standard Model at non-zero Hubble rate, and comparing these to a simpler polynomial model, as well the same Standard Model effective potential with a fixed de Sitter background. Section \ref{sec:bounces} will describe the basics of the CdL prescription for computing tunneling rates. In section \ref{sec:HMeigen} we will discuss the eigenvalue spectrum of the Hawking-Moss instanton and review how this leads to a `critical Hubble rate', $H_{0\rm{crit}}$, below which the Hawking-Moss solution does not contribute to tunneling. In section \ref{sec:numerical} we will discuss the numerical techniques used to find the bounce solutions. This involves numerical challenges arising from the fact that we wish to compute `thick-wall' bounces without using the fixed background approximation, in order to take into account the effects of gravitational backreaction, which is potentially significant since the energy scale of the Standard Model true vacuum is expected to be much larger than the barrier scale or most conceivable inflationary energy scales. In sections \ref{sec:poly} and \ref{sec:SM} we will present the results of calculations in the polynomial model potential, and the Standard Model effective potential, respectively, comparing the calculations with full back-reaction to the results using a fixed de Sitter background for the Standard Model case (section \ref{sec:SM}). We will see that there are significant qualitative differences between the two scenarios, with the polynomial model resulting in a `well-behaved' smooth transition to Hawking-Moss dominance when $H_0$ is raised past the critical threshold, and the Standard Model exhibiting a much sharper transition, with the appearance of additional non-oscillating solutions. Finally, we will discuss what happens as the Hubble rate tends to zero, presenting analytic and numerical arguments that the results should smoothly approach those obtained for a flat false vacuum.

\section{Vacuum decay in de Sitter space: CdL Prescription\label{sec:bounces}}
\subsection{CdL Basics\label{sec:basics}}
The decay rate of a vacuum set by a single scalar field coupled to gravity is given by\cite{PhysRevD.21.3305}:
\begin{equation}
\Gamma = A\exp(-B),
\end{equation}
where $A$ is a prefactor determined by computing functional determinant fluctuations around a bounce solution to the Euclidean action:
\begin{equation}
S_E[\phi,g_{\mu\nu}] = \int\dd^4x \sqrt{|\det g|}\left[\frac{1}{2}\nabla_{\mu}\phi\nabla^{\mu}\phi + V(\phi) - \frac{M_{\rm{P}}^2}{2}R\right].\label{eq:EuclideanAction}
\end{equation}
Throughout this paper, $M_{\rm{P}}$ denotes the `reduced Planck Mass', $M_{\rm{P}} = \frac{1}{\sqrt{8\pi G_{\rm{N}}}}$. The exponent, $B$, is given by the difference between the action of this bounce and the action of the `false vacuum solution', where the field sits in the false vacuum $\phi(\chi) = \phi_{\rm{fv}}$.
\begin{equation}
B = S[\phi,g_{\mu\nu}] -  S[\phi_{\rm{fv}},g_{\rm{fv}\mu\nu}].\label{eq:decayExponent}
\end{equation}
The bounce solution which determines the decay rate is the solution for which $B$ (and thus $S[\phi,g_{\mu\nu}]$) is smallest. Other solutions can contribute to the decay rate, but if their action is larger then they give exponentially suppressed contributions.\\
The smallest action solutions can be found by extremizing the action and solving the Euclidean equations of motion for the coupled gravitational and scalar field. To simplify this calculation, it can be assumed that the dominant solutions are $O(4)$ symmetric. This was proven in the absence of gravity (see \cite{ColemanGlaserMartin1978}), and is believed to be likely when gravity is included and the background respects this symmetry (see references \cite{Garriga:2004nm,Masoumi:2012yy} for a discussion). Under this assumption, the metric can be placed in a co-ordinate system that takes the form:
\begin{equation}
\dd s^2 = \dd\chi^2 + a^2(\chi)\dd\Omega_3^2,\label{eq:O4metric}
\end{equation}
where $\dd\Omega_3^2$ is the metric of a 3-sphere. The $\chi$ co-ordinate is a radial parameter, and $a(\chi)$ describes the radius of curvature of a 3-sphere of co-ordinate radius $\chi$. The equations of motion in this case are:
\begin{align}
\ddot{\phi}& + \frac{3\dot{a}}{a}\dot{\phi} - V'(\phi) = 0\label{eq:scalarField}\\
\dot{a}^2 &= 1 - \frac{a^2}{3M_{\rm{P}}^2}\left(-\frac{\dot{\phi}^2}{2} + V(\phi)\right).\label{eq:scaleFactor}\\
\ddot{a} &= -\frac{a}{3M_{\rm{P}}^2}\left(\dot{\phi}^2 + V(\phi)\right)\label{eq:scaleFactorAcceleration}
\end{align}
Equation (\ref{eq:scaleFactorAcceleration}) is equivalent to differentiating Eq. (\ref{eq:scaleFactor}), but we include it because it is in fact easier to use numerically, due to not requiring a choice of sign when taking the square root of the RHS of Eq. (\ref{eq:scaleFactor}). This is important because in a de Sitter background, $\dot{a}$ does in fact change sign. Note that the first term on the RHS of Eq. (\ref{eq:scaleFactor}) is always 1 is due to the fact that the surfaces of constant $\chi$ always have the geometry of a 3-sphere due to O(4) symmetry, and thus always have positive `spatial' curvature. This does not imply that the geometry of the full four dimensional space is positive, however, and it can in fact be negative in regions where the potential is negative (such as the interior of a nucleated vacuum bubble):
\begin{equation}
R = \frac{\dot{\phi}^2 + 4V(\phi)}{M_{\rm{P}}^2} = \frac{6(1-\dot{a}^2)}{a^2} - \frac{6\ddot{a}}{a}.
\end{equation}
When evaluated at a solution of these equations of motion, the decay exponent take the form:
\begin{equation}
B = \frac{24\pi^2 M_{\rm{P}}^4}{V_0} - 2\pi^2\int_{0}^{\chi_{\rm{max}}}\dd\chi a^3(\chi)V(\phi(\chi)),\label{eq:B_at_sol}
\end{equation}
where $V_0 \equiv V(\phi_{\rm{fv}})$, and $\chi_{\rm{max}}$ is the (possibly infinite) maximum value of $\chi$ that covers the entire patch of the manifold described by this co-ordinate system. Note that throughout this paper we will split the potential as $V(\phi) = V_0 + \Delta V(\phi)$ where $V_0 = V(\phi_{\rm{fv}})$ is the value of the potential in the false vacuum (giving an effective cosmological constant) and $\Delta V(\phi)$ is the rest of the potential, shifted so that $\Delta V(\phi_{\rm{fv}}) = 0$. The effects of varying the Hubble rate are then included by varying $V_0$ and leaving $\Delta V(\phi)$ unchanged. In principle, however, this neglects the fact that changing the background Hubble rate would affect the scale, $\mu$, at which we should evaluate the running couplings (see \cite{PhysRevLett.113.211102} for example) - taking this into account would produce a different scalar potential, which is potentially an important effect of a de Sitter background. We will not consider the effect of that here, and instead simply consider the flat-space Standard Model potential.
\\
To ensure that $B$ is finite, the following boundary conditions are imposed: (1) $a(0) = 0$ (this defines $\chi = 0$ to be the center of the bounce), (2) $\dot{\phi}(0) = 0$ (to guarantee smoothness of the solution at the $a = 0$ co-ordinate singularity) and a final condition (3) which depends on the large $\chi$ behavior of $a(\chi)$. If $a(\chi\rightarrow \infty)\rightarrow \infty$ (or a non-zero constant) then the domain of $\chi$ is infinite and the space non-compact. In that case, $a^3(\chi)V(\phi(\chi))$ must approach zero sufficiently fast that $B$ remains finite. Alternatively, if $\exists$ finite $\chi_{\rm{max}}$ such that $a(\chi_{\rm{max}}) = 0$, then the manifold is compact and we require $\dot{\phi}(\chi_{\rm{max}}) = 0$ to ensure smoothness at this second co-ordinate singularity.
\\
It is straightforward to see that if $V_0 > 0$, the former case cannot have finite action - for $V(\phi)$ to approach zero, $\phi$ must approach a zero of the potential, which is not in general a stationary point of the potential, so the solution will not stay there. When $V_0 = 0$, then the false vacuum is a zero (and a stationary point) - in that case, the former holds: $a(\chi\rightarrow\infty)\rightarrow\infty$ and $\phi(\chi\rightarrow\infty)\rightarrow \phi_{\rm{fv}}$ sufficiently fast that the action is finite (there is an additional complication in that the false vacuum action appears to be infinite in the $V_0\rightarrow 0$ limit, which is addressed in section \ref{sec:V0to0}). As a consequence of this, for $V_0 > 0$, finite action solutions will fall into the latter category, and the boundary conditions can be summarized as $a(0) = 0, \dot{\phi}(0) = \dot{\phi}(\chi_{\rm{max}}) = 0$ where $\chi_{\rm{max}} > 0$ is defined by $a(\chi_{\rm{max}}) = 0$.

\subsection{Types of Solutions\label{sec:types}}
\subsubsection{Hawking-Moss Solution\label{sec:HM}}
The simplest solution to Eqs. (\ref{eq:scalarField}) - (\ref{eq:scaleFactorAcceleration}) with these boundary conditions is the Hawking-Moss solution\cite{HAWKING198235}, which is a constant at the top of the barrier:
\begin{align}
\phi(\chi) =& \phi_{\rm{HM}},\\
a(\chi) =& \frac{\sin(H_{\rm{HM}}\chi)}{H_{\rm{HM}}},
\end{align}
where $H_{\rm{HM}}^2 = \frac{V(\phi_{\rm{HM}})}{3M_{\rm{P}}^2}$. For this the decay exponent is:
\begin{equation}
B = 24\pi^2M_{\rm{P}}^4\left(\frac{1}{V_0} - \frac{1}{V(\phi_{\rm{HM}})}\right).\label{eq:B_HM}
\end{equation}
In the limit where $|V(\phi_{\rm{HM}}) - V_0| \ll |V_0|$, then, this is given approximately by:
\begin{equation}
B\approx \frac{8\pi^2 \Delta V(\phi)}{3H_{\rm{HM}}^4},
\end{equation}
which is the ratio of the energy requires to excite a sphere of radius $\frac{1}{H_{\rm{HM}}}$ to the top of the barrier, over the Gibbons-Hawking temperature $\frac{H_{\rm{HM}}}{2\pi}$. This motivates a thermal interpretation of the Hawking-Moss solution\cite{PhysRevD.76.064003}.\\
\subsubsection{Coleman de Luccia Solution}
There are also non-trivial $O(4)$-symmetric solutions to the equations of motion, which may or may not exist depending on the shape of the potential. Those non-trivial solutions which cross the barrier once, and are monotonic between $\phi(0)$ and $\phi(\chi_{\rm{max}})$ are known as Coleman-de Luccia (CdL) bounces\cite{PhysRevD.21.3305}. Such solutions can be found by the overshoot/undershoot method, which is described in the next section. Note that there need not be only a single CdL solution - Weinberg and Hackworth\cite{PhysRevD.71.044014} found examples of potentials with particularly `flat' barriers (as determined by the ratio $\beta = \frac{|V''(\phi_{\rm{HM}})|}{H_0^2}$) admitting multiple CdL type bounces crossing the barrier only once. The existence of these multiple bounces in the Standard Model is the subject of this paper, and we emphasize that these are not the same as oscillating solutions (discussed below).
\\
\subsubsection{Oscillating Solutions\label{sec:oscillating}}
It is also possible to consider solutions which cross the barrier more than once before settling down to $\dot{\phi}(\chi_{\rm{max}}) = 0$, and these have been investigated by various authors (see for example \cite{PhysRevD.71.044014}). There is strong evidence, however, that these oscillating solutions possess multiple negative eigenvalues in their spectrum of linear fluctuations \cite{PhysRevD.73.083513,PhysRevD.86.124001}: in particular, a bounce crossing the barrier $N$ times has exactly $N$ negative modes. This leads to questions about their relevance for tunneling - Coleman originally argued that bounces with more than one negative eigenvalue in their fluctuation spectrum do not contribute to tunneling, since they correspond to stationary points of the action which are not minima of the set of tunneling paths through the barrier\cite{COLEMAN1988178}. However, this argument comes with the caveat that it does not directly apply to case of tunneling in de Sitter space. Nevertheless, by rephrasing the de Sitter tunneling process in terms of thermally assisted tunneling, Brown and Weinberg argued that this same restriction also applied to de Sitter bounces\cite{PhysRevD.76.064003}. Thus, oscillating solutions should not be regarded as contributing the the tunneling rate. CdL solutions, with a single negative eigenvalue, do contribute to the decay rate.

\section{The critical threshold: eigenvalues of the Hawking-Moss instanton\label{sec:HMeigen}}
In this section we will explain the origins of the `critical threshold' which determines whether Hawking-Moss solutions contribute to vacuum decay. The first step is to understand the behavior of `non-instanton' solutions to Eqs. (\ref{eq:scalarField}) and (\ref{eq:scaleFactor}), that is, solutions which do not satisfy the bounce boundary conditions.

\subsection{Overshoot/Undershoots Solutions\label{sec:noninstanton}}
Most values of $\phi(0)$ will not lead to solutions of the equations of motion that satisfy the bounce boundary conditions, and in fact result in divergent solutions. These `non-instanton' solutions can be categorized into two types, with a precise definition given by Balek and Demetrian\cite{PhysRevD.69.063518}: overshoot solutions are those that diverge on the opposite side of the barrier to that on which they start, and undershoots diverge on the same side. These solutions can then be further categorized by their `order', i.e. the  number of times, $N$, that they cross the barrier - Balek and Demetrian prove that between $\phi_0$ for a non-instanton solution of order $N$ and $\phi_0$ for a non-instanton solution of order $N+1$ there must always lie a bounce solution that crosses $N$ times.
\\
Since we are only interested in the $N = 1$, CdL, solutions in this paper, we adopt the slightly different definition that `undershoot' solutions are those that cross $\dot{\phi} = 0$ before encountering the second $a = 0$ singularity (note that such undershoots could also conceivably be `oscillating' bounce solutions if they then go on to possess another zero of $\dot{\phi}$ coinciding with the $a = 0$ singularity, but since we are only interested in CdL bounces in this paper, we classify these as undershoots too). Solutions which encounter the $a = 0$ coordinate singularity without ever encountering $\dot{\phi} = 0$ are classified as `overshoot' solutions. Using this definition singles out the CdL type solutions, while ignoring the oscillating solutions.

\subsection{Eigenvalues of the Hawking-Moss Solution\label{subsec:HMeigen}}
An attractive feature of the Hawking-Moss solution is that it is possible to compute the spectrum of eigenvalues for linearized field-space fluctuations around it analytically. The \emph{scalar} fluctuations satisfy an equation determined by the second functional derivative of the action:
\begin{equation}
-\nabla_{\mu}\nabla_{\mu}\delta\phi + V''(\phi_{\rm{HM}})\delta\phi = 0,
\end{equation}
where $\nabla_{\mu}\nabla_{\mu}$ is fixed in the constant 4-sphere background of the Hawking-Moss solution. Note that in principle one should consider metric fluctuations as well. However, it is always possible to choose a gauge in which only the scalar fluctuations are relevant for computing the eigenvalue spectrum\cite{PhysRevD.62.083501,PhysRevD.63.123514}. The solutions to this are 4-sphere spherical harmonics (Gegenbauer functions) with eigenvalues:
\begin{equation}
\lambda_{N} = -V''(\phi_{\rm{HM}}) + N(N + 3)H_{\rm{HM}}^2.\label{eq:eigenspectrum}
\end{equation}
As with all bounces, the $N=0$ mode is negative - this is what gives an imaginary contribution to the vacuum energy and a resulting vacuum instability. Of interest here is the $N = 1$ mode, which changes sign when:
\begin{equation}
\frac{V_0 + \Delta V(\phi_{\rm{HM}})}{3M_{\rm{P}}^2} - \frac{V''(\phi_{\rm{HM}})}{4} = 0.
\end{equation}
This defines a critical false vacuum Hubble rate, $H_{\rm{crit}}$, or critical $V_{0\rm{crit}}$:
\begin{equation}
H_{\rm{crit}}^2 = \frac{V_{0\rm{crit}}}{3M_{\rm{P}}^2} = - \frac{V''(\phi_{\rm{HM}})}{4} - \frac{\Delta V(\phi_{\rm{HM}})}{3M_{\rm{P}}^2},
\end{equation}
below which the Hawking-Moss solution always has multiple negative eigenvalues, and thus is expected not to contribute to tunneling.
\\
This critical threshold is significant for tunneling in de Sitter space because it appears to be a value for which there is qualitative change in the behavior of the non-trivial (CdL) solutions. It has been discussed as a possible bound for the existence of CdL bounces in the form of the condition $\beta > 4$, where $\beta = |V''(\phi_{\rm{HM}})|/H_0^2$ for CdL bounces to exist\cite{Jensen1984176}, however, it's actual role is somewhat weaker than this\cite{PhysRevD.69.063518}, and in fact CdL solutions have been found in potentials violating it\cite{PhysRevD.71.044014}.
\\
The boundary conditions mentioned in the previous section describe a two point boundary value problem for the scalar field: $\dot{\phi}(0) = \dot{\phi}(\chi_{\rm{max}}) = 0$, which can be solved by shooting. The procedure is to pick a value of $\phi(0) = \phi_0$, and classify the solution as (1) an `undershoot' or `overshoot', as discussed in section \ref{sec:noninstanton}. Balek and Demetrian\cite{PhysRevD.69.063518} show that between $\phi_0$ leading to an undershoot and $\phi_0$ leading to an overshoot, there must always lie some $\phi_0$ which leads to a bounce solution, by continuity. It is always possible to argue that $\phi_0$ starting sufficiently close to the false vacuum leads to an overshoot (see appendix \ref{sec:fvovershoots}); Balek and Demetrian showed that one can establish the existence of an undershoot for $\phi_0$ sufficiently close to the \emph{barrier}, if $V_0 < V_{0\rm{crit}}$. This implies a CdL bounce \emph{must} exist for $V_0 < V_{0\rm{crit}}$. Above this threshold, existence is not guaranteed, but also not ruled out. Note that it is possible to have no CdL solutions at all: Balek and Demetrian showed that $V(\phi) < -\frac{3M_{\rm{P}}^2V''(\phi)}{4}$ for \emph{some} $\phi$ in the barrier is a necessary (but not sufficient) condition for the existence of a bounce.\\
We will summarize Balek and Demetrian's argument here, as it is pertinent to understanding the role of $V_{0\rm{crit}}$. Consider a solution with $\phi_0$ arbitrarily close to $\phi_{\rm{HM}}$, such that the scalar field equation can be treated as approximately linear, and $a(\chi) \approx \frac{\sin(H_{\rm{}HM}\chi)}{H_{\rm{HM}}}$. Then the scalar field satisfies:
\begin{equation}
\Delta\ddot{\phi} + 3H_{\rm{HM}}\cot(H_{\rm{HM}}\chi)\Delta\dot{\phi} - V''(\phi_{\rm{HM}})\Delta\phi = 0,\label{eq:deltaphi}
\end{equation}
where $\Delta\phi = \phi - \phi_{\rm{HM}}$. The transformation $u = \cos(H_{\rm{HM}}\chi)$ turns this into Gegenbauer's differential equation, and this fact can also be used to derive the eigenspectrum of Eq. (\ref{eq:eigenspectrum}). For the case at hand, however, the Gegenbauer functions for a given $\Delta\phi_0$ can be expressed in terms of the hypergeometric function:
\begin{align}
\Delta\phi(\chi) =& \Delta\phi_0\text{ }{}_2F_1\left(\frac{3}{2} + \alpha,\frac{3}{2} - \alpha,2,\sin^2\left(\frac{H_{\rm{HM}}\chi}{2}\right)\right).\\
\alpha =& \sqrt{\frac{9}{4} - \frac{V''(\phi_{\rm{HM}})}{H_{\rm{HM}}^2}}.\nonumber
\end{align}
Using standard identities for the hypergeometric function, the solution near the second co-ordinate singularity at $\chi_{\rm{max}} = \frac{\pi}{H_{\rm{HM}}}$ is, asymptotically:
\begin{equation}
\Delta\phi(\chi) \sim -\frac{4\Delta\phi_0 \cos\left(\pi\sqrt{\frac{9}{4} - \frac{V''(\phi_{\rm{HM}})}{H_{\rm{HM}}^2}}\right)}{(2 - \frac{V''(\phi_{\rm{HM}})}{H_{\rm{HM}}^2})\pi(\pi - H_{\rm{HM}}\chi)^2}.
\end{equation}
The nature of this solution (overshoot or undershoot) is determined by the sign with which it diverges relative to the sign of $\Delta\phi_0$, and thus by the sign of the cosine in the numerator. An overshoot will diverge on the opposite side of the barrier to where it starts, thus, $\frac{\Delta\phi}{\Delta\phi_0}$ diverges to negative infinity, while undershoots, which fall back before diverging, diverge as $\frac{\Delta\phi}{\Delta\phi_0}\rightarrow  + \infty$. Consequently, as $V_0$ approaches $V_{0\rm{crit}}$ from below, all the solutions are undershoots, but as it approaches from above, all the solutions are overshoots.\\
This is why CdL solutions are not guaranteed above $V_{0\rm{crit}}$: both a solution arbitrarily close to the false vacuum and a solution arbitrarily close to the top of the barrier are overshoots, so unless there is an undershoot somewhere in between them, all solutions between the false vacuum and the top of the barrier are overshoots and no CdL solution exists. If there \emph{is} an undershoot between the false vacuum and the barrier for $V_0 > V_{0\rm{crit}}$, however, then we are in an unusual situation, because starting with $\phi_0$ at the false vacuum and moving towards the top of the barrier, we must transition at least once to a region of undershoots, and then back to a region of overshoots. \emph{Both} these transitions require a \emph{separate} bounce solution to exist, indicating that there are now at least two CdL-type bounces.\\
We can thus conclude one of two things: (1) there are no undershoots on the interval $(\phi_{\rm{fv}},\phi_{\rm{HM}})$, and since the CdL solution which necessarily exists for $V_0 < V_{0\rm{crit}}$ should vary smoothly with $V_0$, we conclude it must smoothly merge with the Hawking-Moss solution as $V_0\rightarrow V_{0\rm{crit}}$ from below, or (2), there are at least two solutions on the interval $(\phi_{\rm{fv}},\phi_{\rm{HM}})$, one of which smoothly merges with the Hawking-Moss solution at $V_{0\rm{crit}}$, and the other does not. The main conclusion of this paper is that the Standard Model effective potential fits into the rather peculiar second category - above $V_{0\rm{crit}}$ there are multiple, distinct, non-oscillating, CdL-like bounce solutions with the same number of turning points.\\
This behavior makes the question of the vacuum decay rate far from simple, as it implies that pairs of bounce solutions can emerge and disappear as $V_0$ is varied, making it difficult to prove that one has found the lowest action solution for a given $V_0$. Indeed, finding \emph{all} the solutions becomes an extremely difficult task, because wide ranges of $\phi_0$ which appear to be all overshoots or all undershoots when sampled can (and, as we will show, do) contain hidden narrow regions of solutions with the opposite character, and associated bounce solutions which are easily missed by a cursory scan.

\section{Numerical Methods\label{sec:numerical}}
The method of finding bounce solutions chosen was the overshoot/undershoot technique proposed by Coleman\cite{PhysRevD.16.1762}. This is a form of non-linear shooting, which consists of picking a value of $\phi_0$ (which is left unspecified by the boundary conditions) and checking whether the solution is an overshoot or an undershoot (see section \ref{sec:noninstanton} for a definition). As discussed in the previous section, in curved space it can be shown that between an overshoot and undershoot solution there always exists a bounce solution\cite{PhysRevD.69.063518}: thus bounce solutions can be found by bisection. This approach is chosen both for its simplicity of implementation, and for the fact that insight into the nature of the solutions can be gained through `scan-plots' like fig. \ref{fig:comb_scanplot}.
\\
A non-trivial feature of the solutions is the fact that the initial conditions are specified at the $a(0) = 0$ co-ordinate singularity. This can be dealt with via a Taylor expansion to a small radial coordinate $\chi$, easily derived from the equations of motion:
\begin{align}
\phi(\chi) \approx & \phi_0 + \frac{V'(\phi_0)}{8}\chi^2,\label{eq:phiboundary}\\
a(\chi) \approx & \chi - \frac{V(\phi_0)\chi^3}{18M_{\rm{P}}^2}\label{eq:aboundary}.
\end{align}
There is a specific feature of the overshoot-undershoot procedure that is unique to the case of de Sitter bounces: in flat space, and in the $V_0 = 0$ case, the bounce is in a sense `one-sided' because it satisfies the boundary condition $\phi(\chi\rightarrow\infty)\rightarrow \phi_{\rm{fv}}$. In the de-Sitter case, however, the condition $\dot{\phi}(\chi_{\rm{max}}) = 0$ implies that there is another unknown parameter, $\phi_{\rm{end}} = \phi(\chi_{\rm{max}})$. This can be found in a similar way to $\phi_0$ simply by applying the overshoot-undershoot procedure on the false vacuum side of the barrier (since $\phi_{\rm{end}}$ must lie in the range $\phi_{\rm{fv}} \leq \phi_{\rm{end}} \leq \phi_{\rm{HM}}$). The two sides of the solution can then be patched together to obtain the entire solution. We will denote these two halves of the solution as the `true-vacuum side' solution and the `false-vacuum side' solution. The false vacuum side must be flipped by the transformation $\chi\rightarrow \chi_{\rm{max}} - \chi$ in order to patch together with the true vacuum side. The matching point at which the solutions meet can be chosen arbitrarily, but we choose it to be the point where $\dot{a} = 0$ (this always exists sufficiently close to a de Sitter bounce, because otherwise the bounce would be non-compact and have infinite action). This procedure has the advantage of avoiding the $a(\chi_{\rm{max}}) = 0$ coordinate singularity when computing the bounce solution, since the solutions are integrated from each side and meet in the middle: integration is always performed out of the singularity (this is important, because integrating into a singularity is numerically unstable, due to the negative friction term in Eq. (\ref{eq:scalarField})  when $\dot{a} < 0$, leading to exponential growth of any small errors).
\\
These methods all apply to finding bounces in general, but there is also a significant challenge which is not present flat space calculations - this is the fact that Eq. (\ref{eq:B_at_sol}) contains a divergent term in the $V_0\rightarrow 0$ limit. If the action is to approach the $V_0 = 0$ result, which is finite, then calculation of the decay exponent, $B$, must involve a cancellation between two large (and ultimately divergent) numbers. This poses a problem for calculations performed at double precision. In the literature, this problem is usually avoided by choosing the fixed background approximation, that is, assuming that $a(\chi)$ takes the same form for the bounce solution as it does in the false vacuum ($a(\chi) = \sin(H_0\chi)/H_0$), which is equivalent to ignoring the effects of back-reaction from the bounce solution on the metric. We choose not to use the fixed background approximation in this paper, because the depth of the Standard Model potential compared to reasonable inflationary scales is large. Consequently we have developed techniques for finding the bounce solutions taking into account all the back-reaction corrections. In particular we do two things: (1) use arbitrary precision numbers, rather than double precision numbers, to perform the calculation (the calculations in this paper use 100 decimal places of precision) and (2) re-write Eq. (\ref{eq:B_at_sol}) in such a way that cancellations of large numbers are avoided where possible.\\
The re-writing of the action we chose is the following: we split it into three parts, $B = B_1 + B_2 + B_3$, where
\begin{align}
B_1 =& -2\pi^2\int_{0}^{\chi_{\rm{max}}}\dd\chi a^3(\chi)\Delta V(\phi(\chi)),\label{eq:B1}\\
B_2 =& -6\pi^2M_{\rm{P}}^2\int_{0}^{\chi_{\rm{max}}}\dd\chi\left[3\sin^2(H_0(\chi_{\rm{max}} - \chi))\delta a_{H_0}(\chi)\right.\nonumber\\
& \left.+ 3H_0\sin(H_0(\chi_{\rm{max}} - \chi))\delta a_{H_0}^2(\chi) + H_0^2\delta a_{H_0}^3(\chi)\right],\label{eq:B2}\\
B_3 =& -\frac{2\pi^2M_{\rm{P}}^2}{H_0^2}(1 + \cos(H_0\chi_{\rm{max}}))^2(\cos(H_0\chi_{\rm{max}}) - 2).\label{eq:B3}
\end{align}
$H_0 = \sqrt{\frac{V_0}{3M_{\rm{P}}^2}}$, and $\delta a_{H_0}(\chi)$ is defined by:
\begin{equation}
a(\chi) = \frac{1}{H_0}\sin(H_0(\chi_{\rm{max}} - \chi)) + \delta a_{H_0}(\chi).\label{eq:asplit}
\end{equation}
In other words, $\delta a_{H_0}(\chi)$ represents the deviation of $a(\chi)$ from the false-vacuum-solution scale factor. Note that there is some freedom here - $\delta a_{H_0}(\chi)$ could have been defined as $a(\chi) = \frac{1}{H_0}\sin(H_0\chi) + \delta a_{H_0}(\chi)$, or any other phase shift of this. However, as we will see in section \ref{sec:V0to0}, Eq. (\ref{eq:asplit}) is the definition that agrees with the deviation of $a(\chi)$ from the false vacuum solution in the $V_0\rightarrow 0$ limit. This makes it the most natural choice.
\\
$B_3$ is an analytic term, and like $B_2$ it arises due to the differing sizes of the bounce solution 4-sphere geometry and the false vacuum 4-sphere. As such, $B_2$ and $B_3$ are both expected to be very small if the fixed background approximation works well. Although it is not obvious, $B_2$ and $B_3$ can be shown to vanish in the $V_0\rightarrow 0$ limit \emph{if} there exists a family of solutions that smoothly approaches the $V_0 = 0$ solution. This is discussed in section \ref{sec:V0to0}.\\
$B_1$ and $B_2$ are evaluated as if they were separate components of the differential equation:
\begin{align}
\frac{\dd B_1}{\dd\chi} =& -2\pi^2 a^3(\chi)V(\phi(\chi))\label{eq:B1eq}\\
\frac{\dd B_2}{\dd\chi} =& - 6\pi^2M_{\rm{P}}^2\left[ 3\sin^2(H_0\chi)\delta a_{H_0}(\chi) +\right. \nonumber\\
& \left.3H_0\sin(H_0\chi)\delta a_{H_0}^2(\chi) + H_0^2\delta a_{H_0}^3(\chi)\right].\label{eq:B2eq}
\end{align}
Notice that Eq. (\ref{eq:B2eq}) differs from Eq. (\ref{eq:B2}) in that it uses $\sin(H_0\chi)$ instead of $\sin(H_0(\chi_{\rm{max}} - \chi))$. This is because $\chi_{\rm{max}}$ is not known a priori until the solution has been computed. However, because we use the method of integrating from both sides, the correct $\delta a_{H_0}(\chi)$ and $\sin(H_0(\chi_{\rm{max}} - \chi))$ terms are obtained when integrating from the false vacuum side of the barrier, for which it is necessary to transform $\chi\rightarrow \chi_{\rm{max}} - \chi$ to patch together with the true vacuum side of the solution. However, the procedure gives the wrong $\delta a$ for the true vacuum half:  the relationship between the correct $\delta a_{H_0}(\chi)$ and the one obtained from the true-vacuum side of the bounce is just a difference of two $\sin$ functions once $\chi_{\rm{max}}$ is determined:
\begin{equation}
\delta a_{H_0}(\chi) = \delta a_{H_0}^{\rm{tv}}(\chi) + \frac{1}{H_0}(\sin(H_0\chi) - \sin(H_0(\chi_{\rm{max}} - \chi))).
\end{equation}
The result of integrating $B_2$ from the true-vacuum side of the barrier up to the matching point is then also simply related to the the contribution it should give (with the correct $\delta a_{H_0}(\chi)$) by a simple analytically calculable function of $\chi_{\rm{max}}$ and the matching point. Thus in principle the fact that $\chi_{\rm{max}}$ is not known a-priori doesn't pose a significant problem, as the contribution to $B_2$ obtained can be easily transformed into the correct contribution. In practice, however, because $\delta a_{H_0}(\chi)$ is frequently small compared to the sinusoidal terms, there are sometimes  situations where doing this leads to significant inaccuracies due to numerical errors in the computed value of $\chi_{\rm{max}}$. Such a situation is fortunately easy to detect because it shows up a discontinuity in $\delta a_{H_0}(\chi)$ when the already correct false-vacuum side of the solution and the corrected true-vacuum side of the solution are patched together. For such situations, it is generally more accurate to compute the whole of $B_2$ using the nearest undershoot solution computed from the false vacuum half of the solution alone, integrating all the way up to $\dot{\phi}(\chi_{\rm{max}}) = 0$ (undershoot solutions are more reliably close to the bounce solution than overshoots, because they can be terminated at the $\dot{\phi} = 0$ point before they diverge, while overshoot solutions are generally not as easy to identify until they have started diverging).

\FloatBarrier
\section{Example - Polynomial  Potential\label{sec:poly}}
\begin{figure}[htb]
\includegraphics[width = 0.5\textwidth]{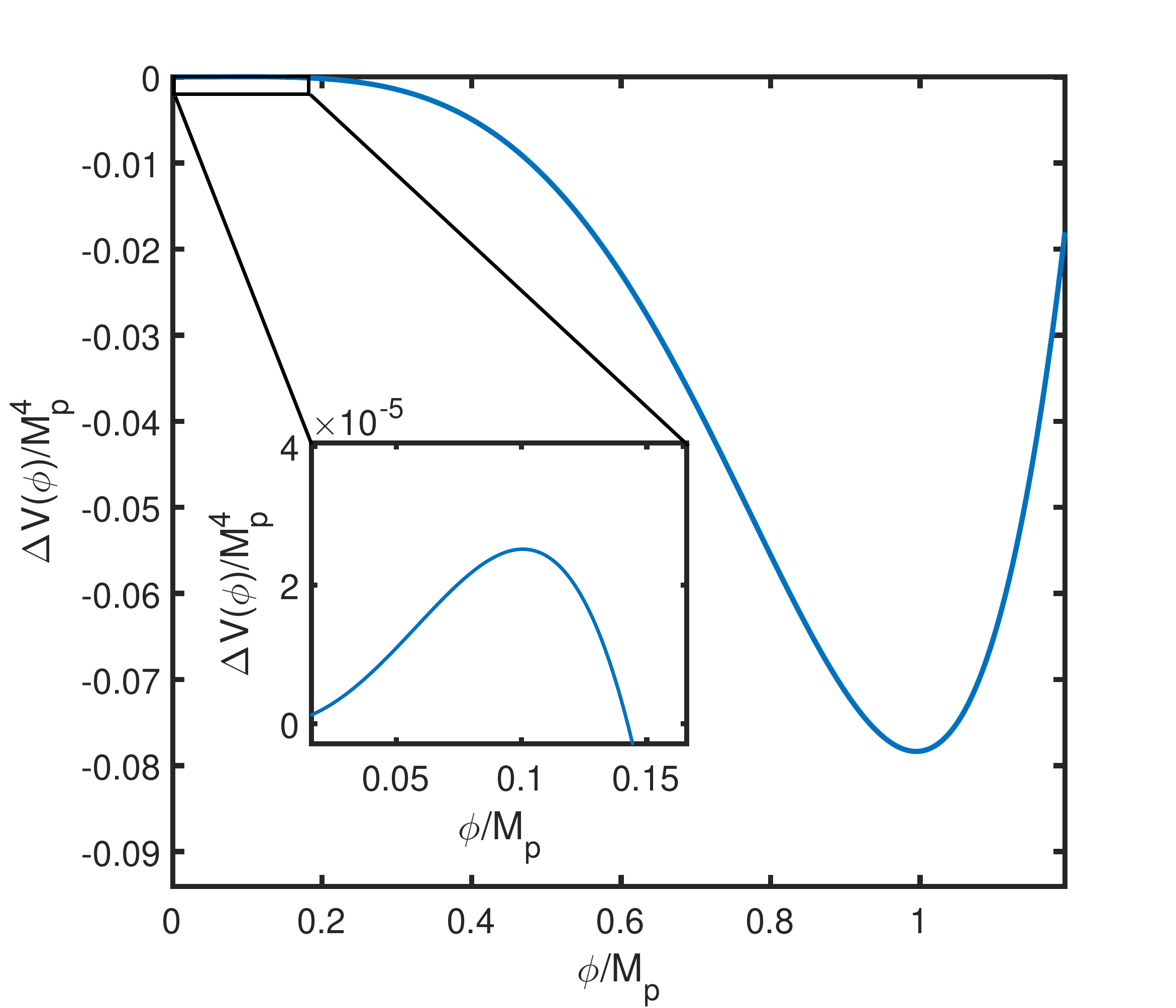}
\caption{\label{fig:poly_pot}Polynomial potential of Eq. (\ref{eq:polypot}) for $\lambda_4 = -1,\lambda_6 = +1,m^2 = 0.1M_{\rm{P}}^2, M = M_{\rm{P}}$.}
\end{figure}

We will first consider an example of de Sitter bounces in a simple polynomial potential.The potential we use is:
\begin{equation}
V(\phi) = V_0 + \frac{1}{2}m^2\phi^2 + \frac{\lambda_4}{4}\phi^4 + \frac{\lambda_6}{6M^2}\phi^6,\label{eq:polypot}
\end{equation}
in this case choosing $\lambda_4 = -1$, $\lambda_6 = +1$, $m^2 = 0.1 M^2$ and $M = M_{\rm{P}}$, which gives the potential in figure \ref{fig:poly_pot}.

\begin{figure*}[p]
\includegraphics[width=\textwidth]{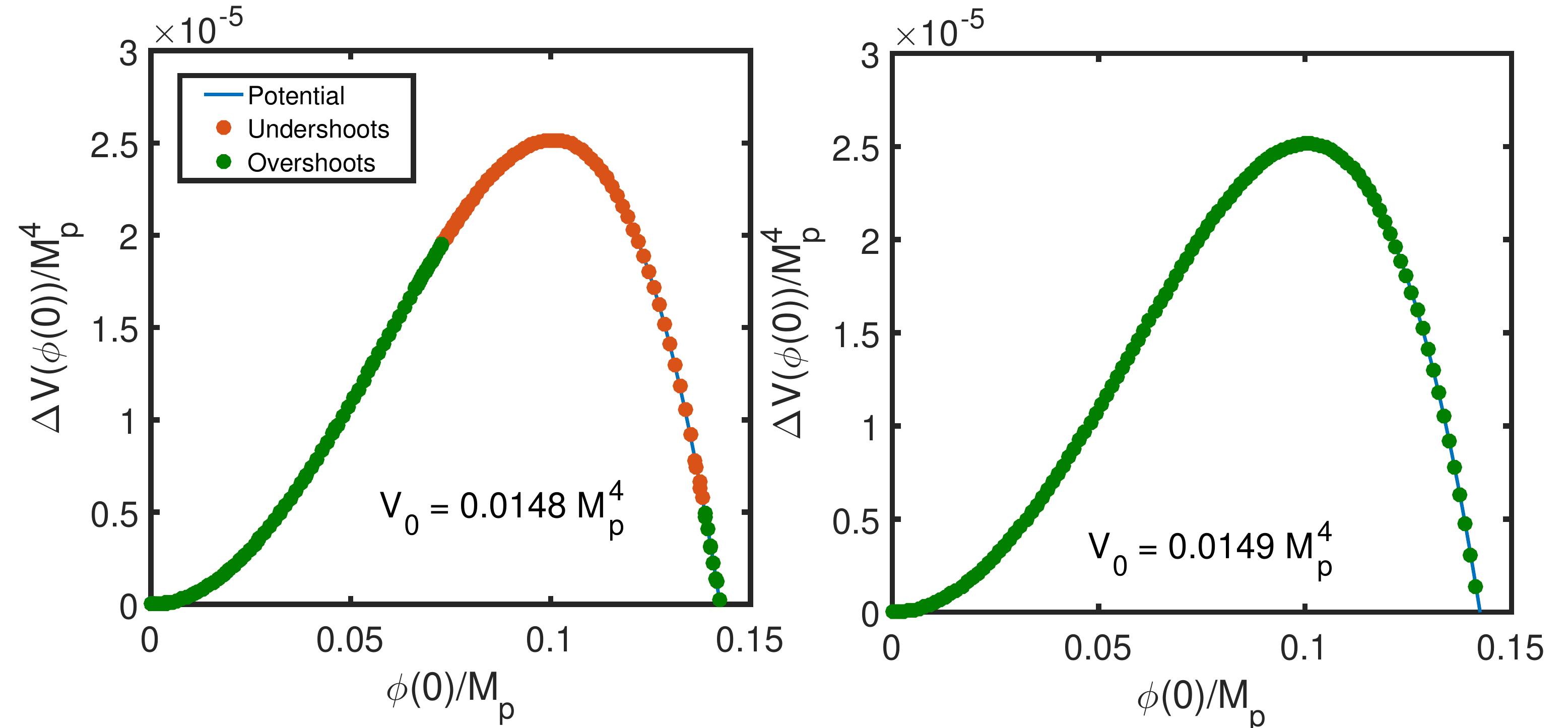}
\caption{\label{fig:poly_over_under}Overshoot/Undershoot structure for the polynomial potential Eq. (\ref{eq:polypot}), above and below the critical threshold at $V_{0\rm{crit}} = 0.01482 M_{\rm{P}}^4$. Below the threshold there are undershoot solutions for $\phi(0)$ sufficiently close to the top of the barrier: above it these solution disappear and all the solutions found are overshoots. This implies that no CdL-type solution exists above $H_{0\rm{crit}}$, and the Hawking-Moss solution is the only contributor.}
\end{figure*}

For these values, the critical Hubble rate is at $V_{0\rm{crit}} = 0.01482 M_{\rm{P}}^4, H_{0\rm{crit}} = 0.0705 M_{\rm{P}}$. In fig. \ref{fig:poly_over_under}, the overshoot/undershoot structure of solutions for various values of $\phi(0)$ is plotted, so as to determine the spectrum of bounce solutions for different $V_0$. This shows the expected behavior; above $V_{0\rm{crit}}$, there are no undershoot solutions at all, and thus no CdL solution exists.

\begin{figure}[htb]
\includegraphics[width=0.5\textwidth]{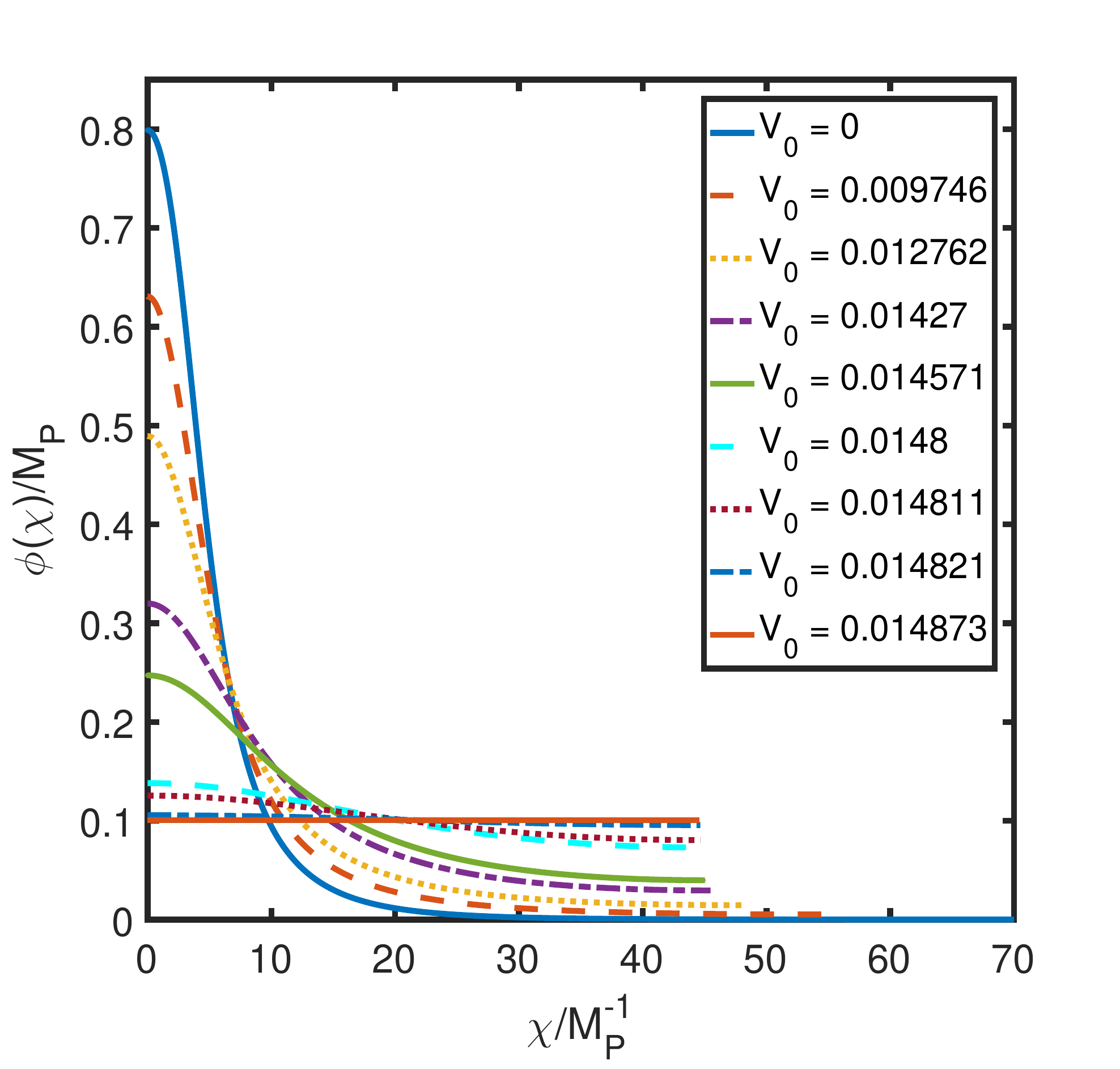}
\caption{\label{fig:poly_sols}Plots of the CdL solutions for various values of $V_0$ in the polynomial potential, showing how they gradually approach the Hawking-Moss solution as $V_0\rightarrow V_{0\rm{crit}}$. For reference, $V_{0\rm{crit}} = 0.014822 M_{\rm{P}}^4$.}
\end{figure}

The bounce solution can then be computed by means of a binary search on the boundaries between overshoot and undershoot regions, as discussed in section \ref{sec:numerical}. Example solutions are shown in fig. \ref{fig:poly_sols}, which shows how the solutions approach the Hawking-Moss solution sitting at the top of the barrier as $V_0$ is raised past the critical value. With increasing $V_0$, the solutions decrease in amplitude, which can be interpreted as thermal effects becoming more and more important compared to quantum tunneling effects\cite{PhysRevD.76.064003}.


\begin{figure*}[p]
\includegraphics[width=\textwidth]{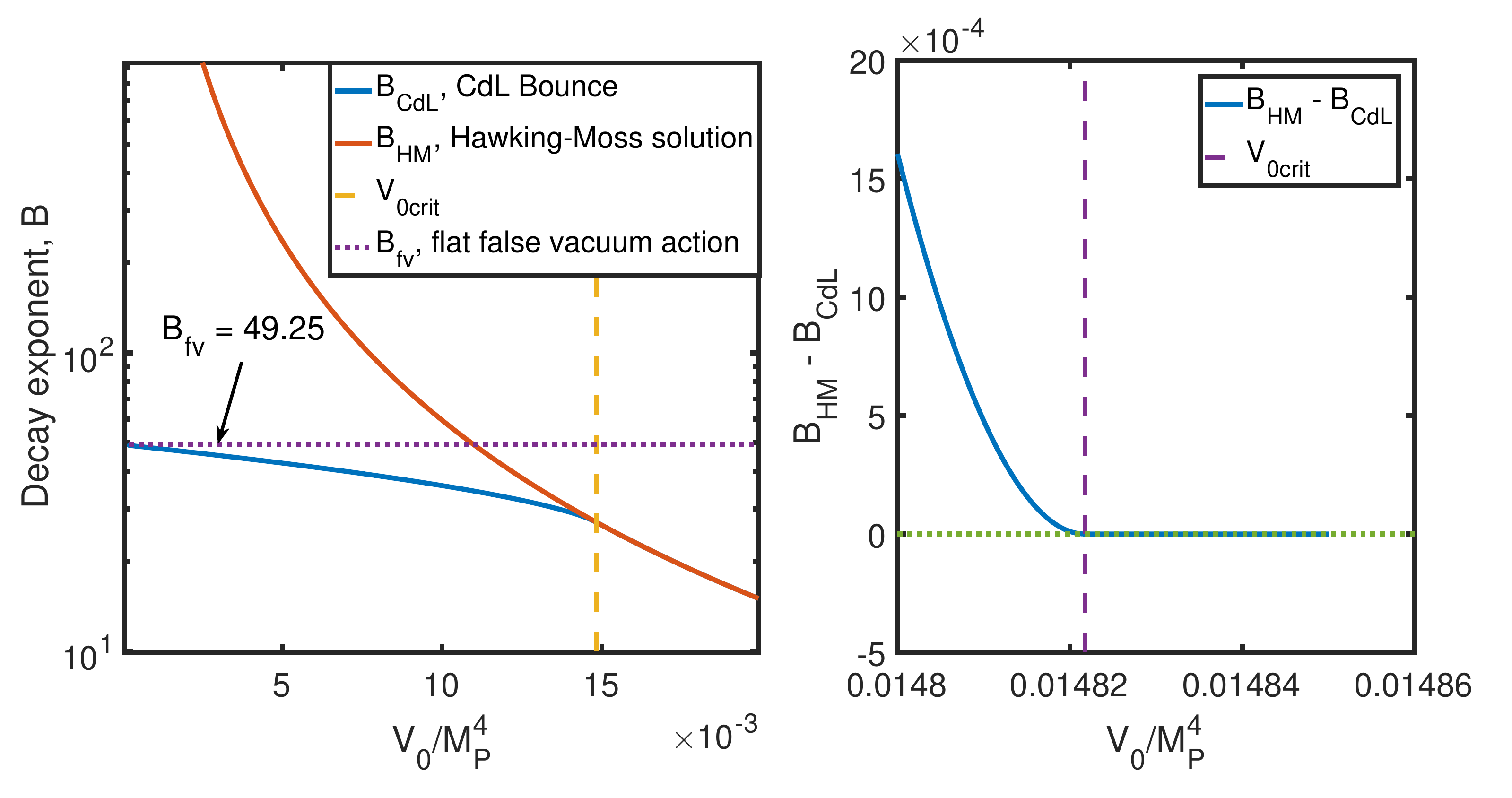}
\caption{\label{fig:poly_action}Left: action of the CdL and Hawking-Moss solutions for the polynomial potential Eq. (\ref{eq:polypot}). The CdL action continuously approaches the flat false vacuum $V_0 = 0$ result below $V_{0\rm{crit}}$, but appears to merge with the Hawking-Moss at the critical threshold. Right: Difference between Hawking-Moss and CdL decay exponents, showing that this falls to zero precisely at the critical threshold.}
\end{figure*}

The resulting decay exponent, $B$, is plotted in figure \ref{fig:poly_action}. This shows a fairly typical behaviour for ``well behaved'' potentials - below the critical threshold, there is a unique CdL bounce whose action approaches the flat false vacuum ($V_0 = 0$) case as $V_0 \rightarrow 0$. At the critical threshold, it appears that the CdL action merges with the Hawking-Moss solution, just as the solutions appear to do in fig. \ref{fig:poly_sols}. 
To check whether $V_{0\rm{crit}}$ really is the point at which the CdL bounce merges with the Hawking-Moss solution, we plot in fig. \ref{fig:poly_action} the difference between the Hawking-Moss and CdL actions for a give $V_0$, in the vicinity of $V_{0\rm{crit}}$. Above $V_{0\rm{crit}}$ the overshoot-undershoot procedure yields the Hawking-Moss solution, because no CdL solution exists, thus the difference is zero. Below $V_{0\rm{crit}}$, it can be seen that the difference between the decay exponents smoothly approaches zero at $V_{0\rm{crit}}$. Note that for $V_0 = V_{0\rm{crit}}$, it is known from linear analysis that the $N=1$ eigenfunction of the Hawking-Moss satisfies the bounce boundary conditions\cite{PhysRevD.69.063518}.

\FloatBarrier
\section{Bounces in the Standard Model\label{sec:SM}}
\begin{figure*}[t!]
\includegraphics[width=\textwidth]{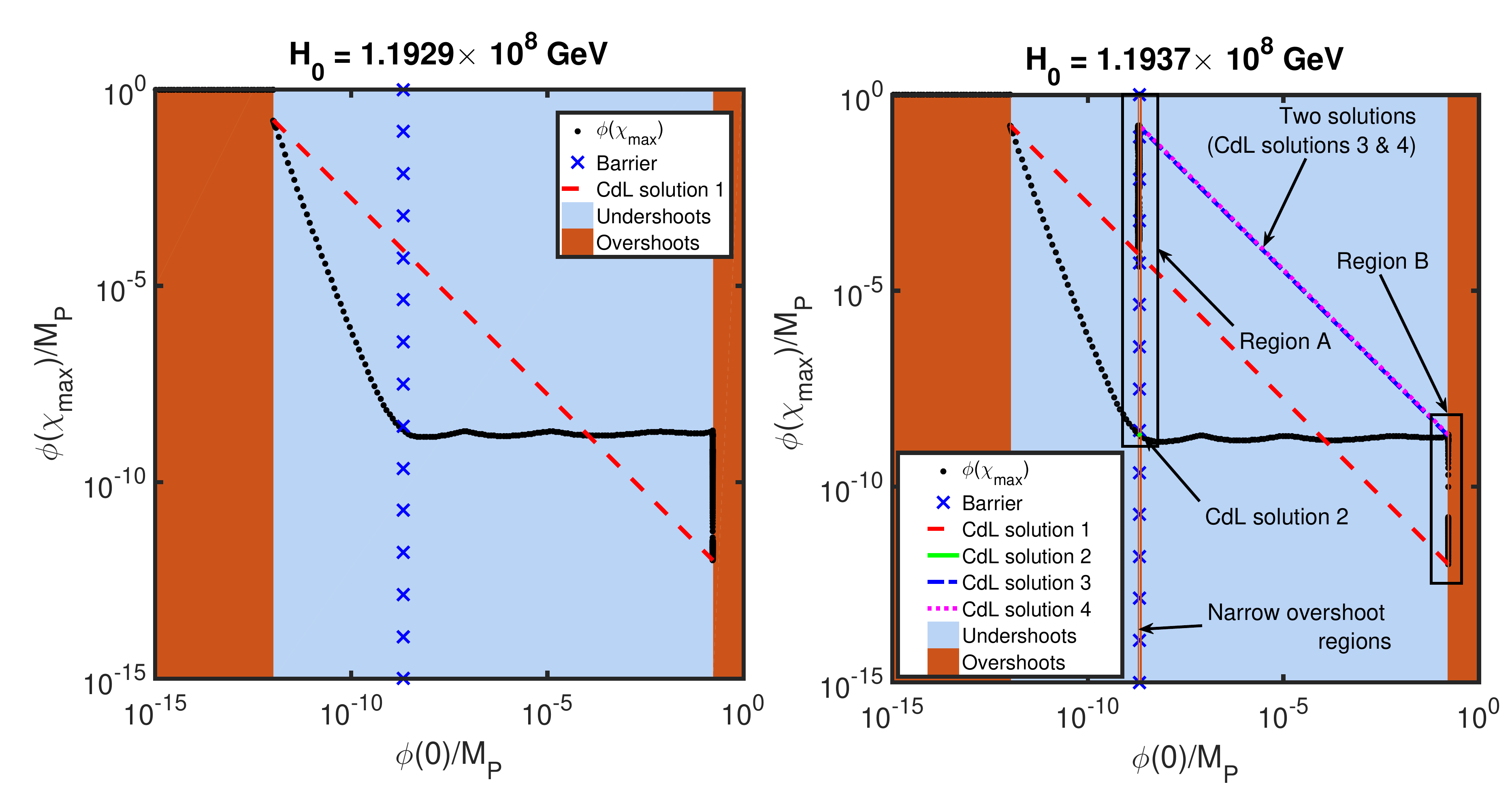}
\caption{\label{fig:comb_scanplot}Scan through possible $\phi(0)$ values in the Standard Model effective potential, giving the resulting $\phi(\chi_{\rm{max}})$ at which $\dot{\phi}(\chi_{\rm{max}}) = 0$ if the solution is an undershoot. Straight lines are drawn between the start and ends of each bounce solution at an overshoot/undershoot boundary. When crossing the critical threshold of $H_{0\rm{crit}} = 1.1931\times 10^8 \rm{ GeV}, (V_{0\rm{crit}} = 7.203\times 10^{-21}M_{\rm{P}}^4)$, there is a dramatic change in the nature of the non-instanton solutions which start close to the barrier (region A, see figure \ref{fig:regionA}), and for those near the top of the CdL bounce (region B, see figure \ref{fig:regionB}).}
\end{figure*}
To study the situation in the Standard Model, consider the following approximation to the effective potential:
\begin{equation}
V(\phi) = \frac{\lambda(\phi)}{4}\phi^4.
\end{equation}
Where $\lambda(\mu)$ is the Higgs self coupling at energy scale $\mu$. In this example we use three loop running of the Standard Model couplings. The potential uses a piecewise polynomial approximation identical to the approach described in a previous paper\cite{PhysRevD.95.025008}. This admittedly has shortcomings - a more correct treatment would modify the scale $\mu$ to include curvature terms\cite{PhysRevLett.113.211102}, but for the purposes of this paper we will consider the flat space potential.\\
\begin{figure}[htb]
\includegraphics[width=0.5\textwidth]{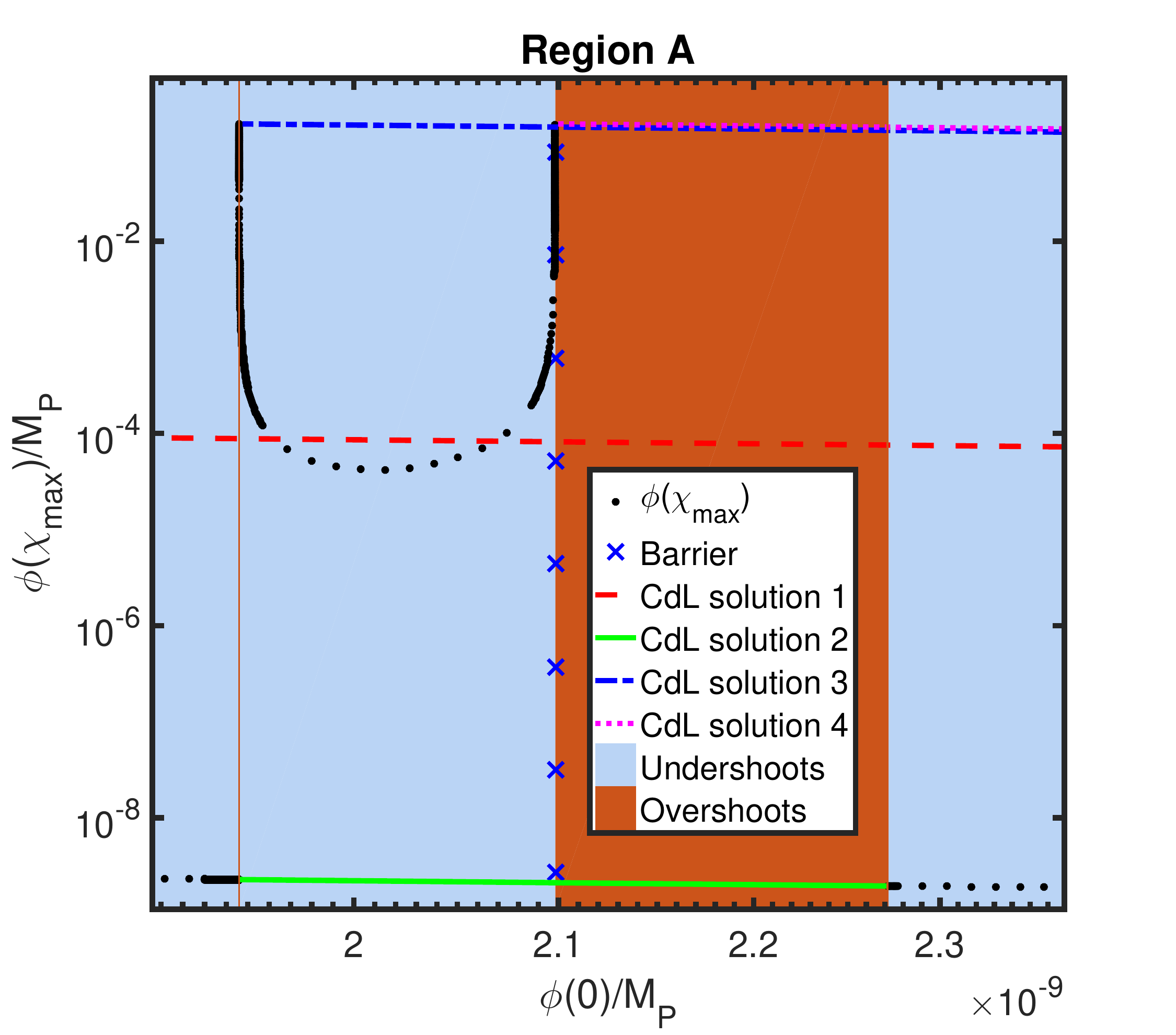}
\caption{\label{fig:regionA}Zoomed scan plot for region A on fig. \ref{fig:comb_scanplot}. An additional narrow overshoot region on the left is exaggerated to improve visibility. }
\end{figure}
\begin{figure}[htb]
\includegraphics[width=0.5\textwidth]{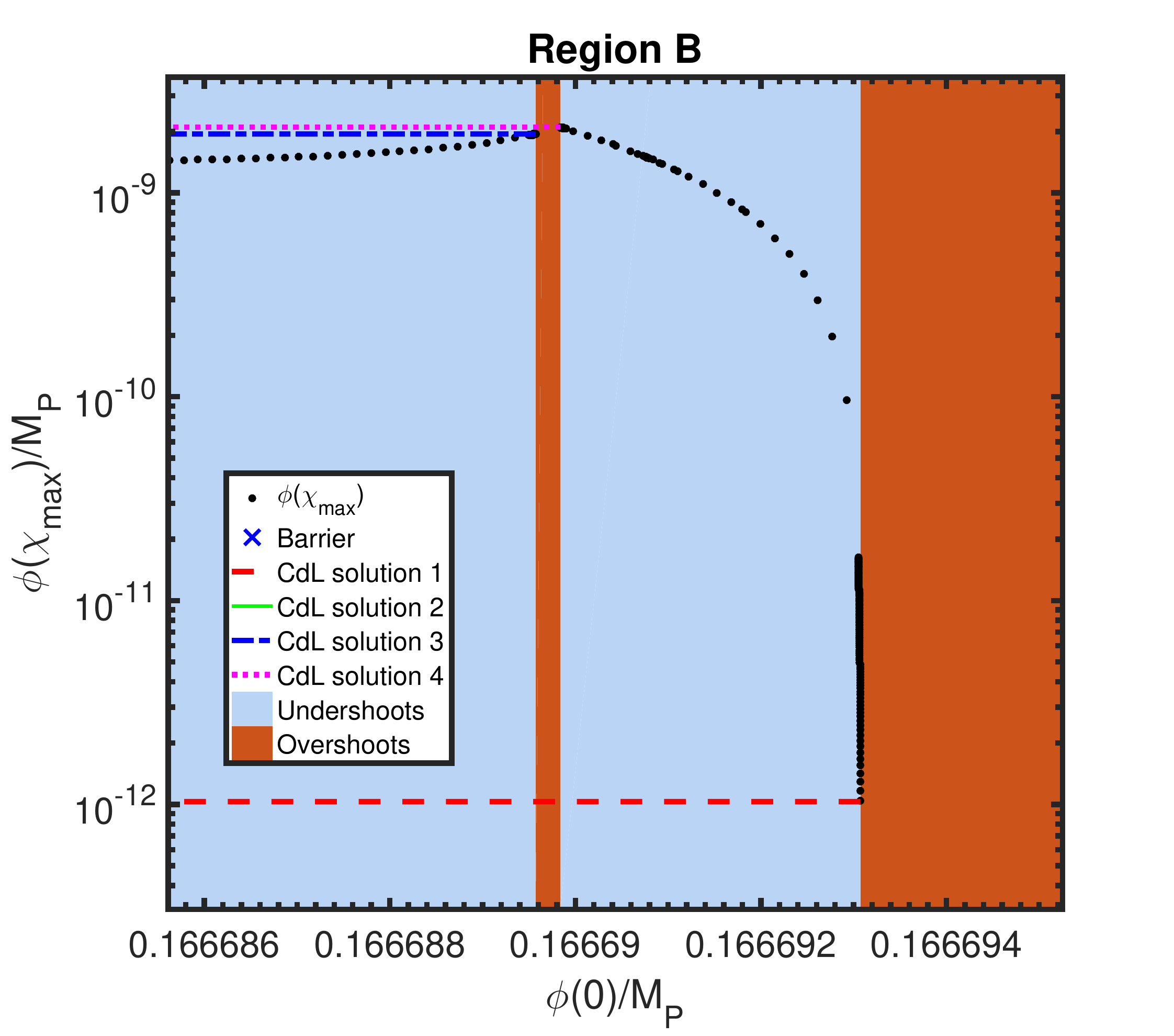}
\caption{\label{fig:regionB}Zoomed scan plot for region B on fig. \ref{fig:comb_scanplot}. Notice the narrow region of overshoots which is not large enough to be visible on fig. \ref{fig:comb_scanplot}.}
\end{figure}
\begin{figure}[htb]
\includegraphics[width=0.5\textwidth]{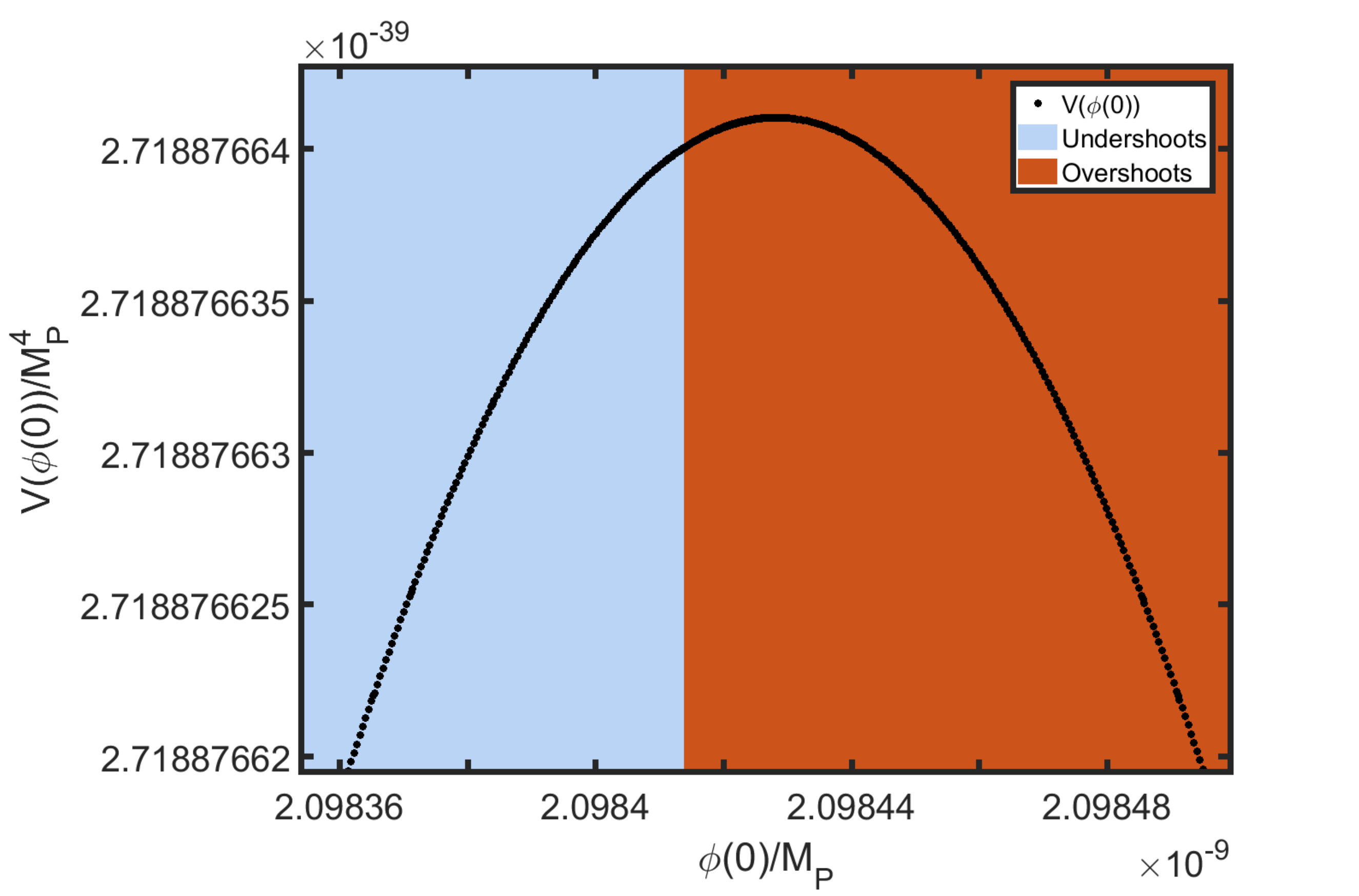}
\caption{\label{fig:bar_top}Plot of the overshoot/undershoot behavior about the top of the barrier of the Standard Model effective potential, for $V_0 = 7.210\times 10^{-21}M_{\rm{P}}^4, H_0 = 1.1937\times 10^8 \rm{ GeV} > H_{0\rm{crit}}$, confirming that solutions sufficiently close to the barrier are overshoots.}
\end{figure}
\begin{figure*}[tb!]
\includegraphics[height=0.35\textheight]{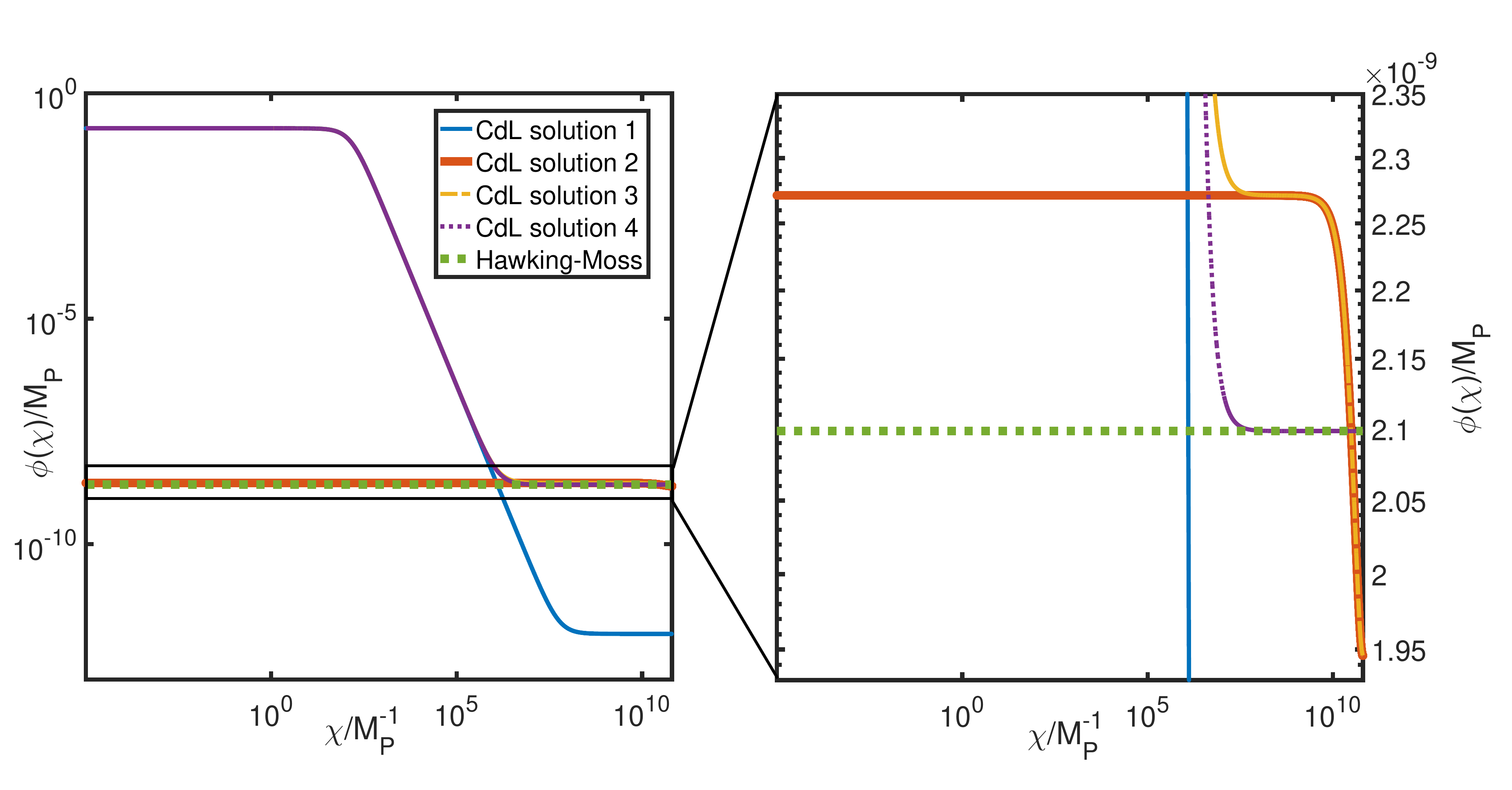}
\caption{\label{fig:sm_bounce_plots}Left: CdL bounce solutions for $V_0 = 7.210\times 10^{-21} M_{P}^{4} (H_0 = 1.1937\times 10^8 \rm{ GeV})$. Three solutions are similar to CdL solution 1 in their interior, but on the false vacuum side they approach the barrier, while the CdL solution 1 reaches much further down. A further solution (CdL solution 2) straddles the barrier with small amplitude. Right: zoomed view of bounce solutions around the barrier. Initial and final values of these solutions are given in table \ref{tab:bounce_data}. CdL solutions are numbered by order of the proximity of their false vacuum end-value to the false vacuum, thus CdL 1 always corresponds to the largest amplitude solution.}
\end{figure*}
\begin{figure*}[htb]
\includegraphics[height=0.35\textheight]{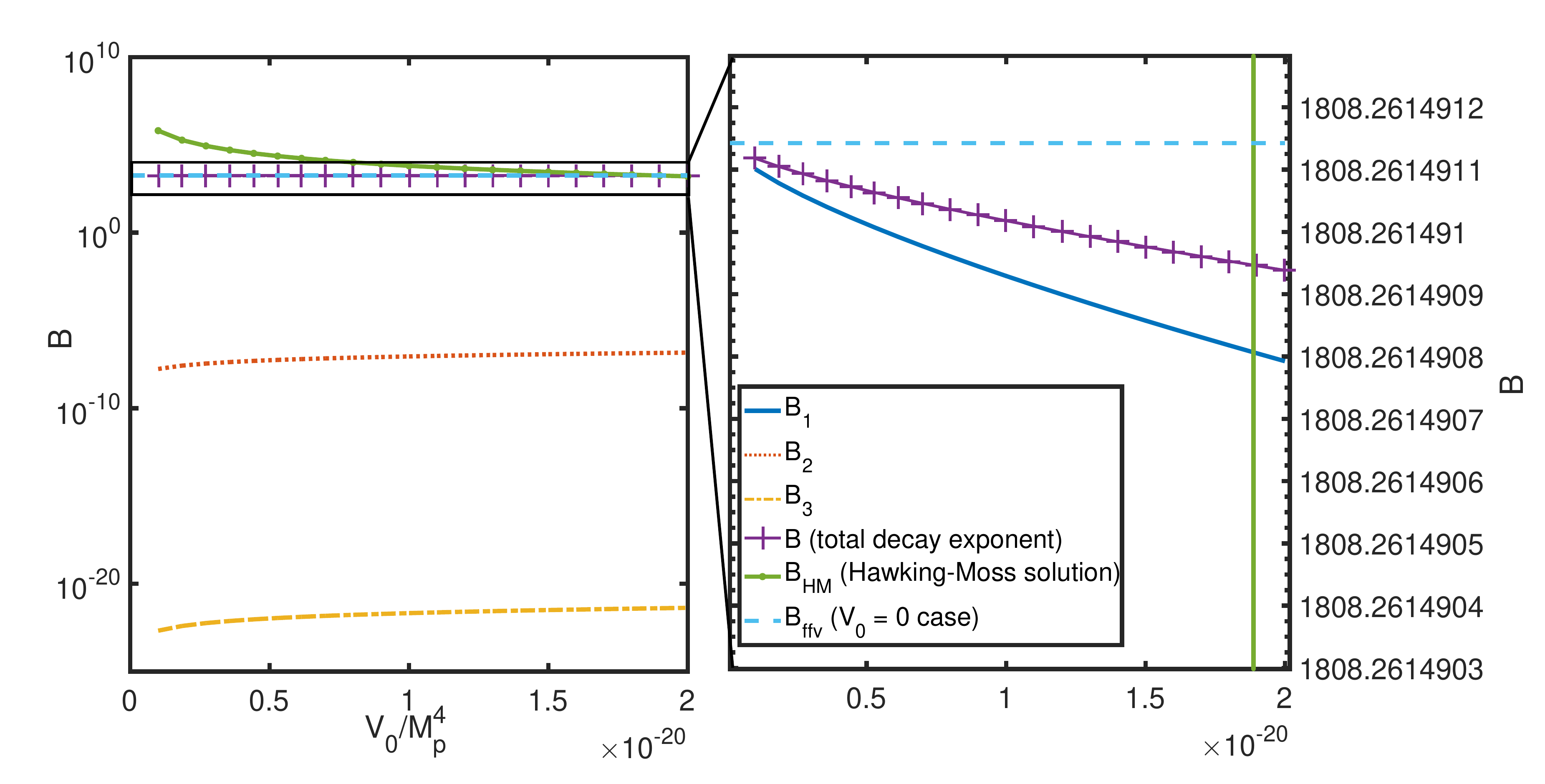}
\caption{\label{fig:action_plot}Plot of the CdL and Hawking-Moss decay exponent, $B$, and its three components $B_1,B_2,B_3$ (see Eqs. (\ref{eq:B1}) - (\ref{eq:B3}) for definitions), for different values of $V_0$, compared to the flat false vacuum ($V_0 = 0$) case.}
\end{figure*}
Performing the numerical calculations with $V_0 > 0$, however, poses significant numerical challenges compared to the $V_0 = 0$ case, which are addressed in the section \ref{sec:numerical}. One issue of physical relevance is the large range of scales in the Standard Model - the behavior of bounces with $V_0 = 0$ is dominated by a scale just below the Planck scale\cite{Isidori2001387}, but for the central values of the Higgs boson and top quark masses, the barrier lies at around $10^{10}\rm{ GeV}$. In this paper we fix $M_h = 125.15 \rm{ GeV}$, $M_t = 173.34 \rm{ GeV}$ and consider only variations of $V_0$. This gives a barrier scale of $\phi_{bar} = 5.110\times 10^9 \rm{ GeV}$, and a critical Hubble rate $H_{0\rm{crit}} = 1.1931\times 10^8 \rm{ GeV}, (V_{0\rm{crit}} = 7.203\times 10^{-21}M_{\rm{P}}^4)$. If the Standard Model behaved like the polynomial potential, therefore, it might be expected that CdL bounces would dominate the decay rate below $H_{0\rm{crit}}$ and Hawking-Moss solutions would dominate above this. In fact, as we will see, the behavior is somewhat different.
\\

First, we plot the overshoot/overshoot structure in logarithmic space (see fig. \ref{fig:comb_scanplot}). 
To understand the structure, we compute the end value $\phi(\chi_{\rm{max}})$ for a given set of points, $\phi(0)$. The `end-value' $\chi_{\rm{max}} > 0$ is defined by either (1) the point where $\dot{\phi} = 0$ if the solution is an undershoot, or (2) the point at which the solution crosses the `overshoot threshold' with positive $\dot{\phi}^2$ if it is an overshoot: either the false vacuum or the true vacuum, depending on which side of the barrier the solution starts on. Note that in the Standard Model the true vacuum is many orders of magnitude larger than the Planck scale, and thus it is necessary for practical reasons to impose a smaller cutoff at which a solution is declared to be an overshoot: for these calculations we chose this to be the Planck scale, but we have checked that the classification is not sensitive to this. The result of this will be a curve which possesses discontinuities at the boundary between a region of undershoots and a region of overshoots. The undershoot solutions, in this case, more closely approximate the true bounce solution because they are terminated before hitting the co-ordinate singularity; their $\phi(\chi_{\rm{max}})$ value is of greatest interest. Figure \ref{fig:comb_scanplot} shows these overshoot/undershoot structures for $H_0 < H_{0\rm{crit}}$ and $H_0 > H_{0\rm{crit}}$. We select these two value of $H_0$ in particular because they are very close to the critical Hubble rate, and illustrate that a dramatic change occurs there. The end $\phi(\chi_{\rm{max}})$ and starting points, $\phi(0)$, of the solutions associated to each overshoot/undershoot transition are plotted as lines interpolating between the different overshoot/undershoot boundary discontinuities.\\
These plots indicate that there is significant structure present, especially above $H_{0\rm{crit}}$. This is the first hint that the Standard Model potential is not a typical potential. For $H_0 < H_{0\rm{crit}}$, the structure is relatively simple; there is a single discontinuity around $\phi(0) \sim 10^{-12}M_p$, and another one at $\phi(0) \sim 0.17 M_p$. Everything in between these points is an undershoot, and everything outside them is an overshoot (note that $\phi(\chi_{max})$ for overshoots through the false vacuum are not shown on these plots because the false vacuum is chosen to be at $\phi = 0$, placing it at $-\infty$ in logarithmic space). As expected, if we define the function $\phi_{\rm{end}}(\phi_0) = \phi_{\phi_0}(\chi_{\rm{max}})$, where $\phi_{\phi_0}(\chi)$ is the solution for a given value of $\phi_0$, then the result is a smooth function between these two discontinuities.\\
The situation for $H_0 > H_{0\rm{crit}}$ is significantly different. We plot in fig. \ref{fig:regionA} the region around the barrier and in fig. \ref{fig:regionB} the region around the top of the CdL bounce, both of which show significant changes.\\

The region around the barrier shows very fine structure, with several more discontinuities in $\phi_{\rm{max}}(\phi_0)$. These discontinuities correspond to narrow regions of $\phi_0$ for which there are overshoots, surrounded by undershoots, and vice-versa. Recall that it was shown analytically that solutions sufficiently close to the barrier are \emph{always} overshoots for $H_0 > H_{0\rm{crit}}$. At first glance, it appears as though all the solutions for $\phi_0 < \phi_{\rm{HM}}$ are undershoots, although this analytic result clearly holds for $\phi_0 > \phi_{\rm{HM}}$. However, a closer inspection reveals that going very close to the barrier does in fact produce overshoots eventually (see figure \ref{fig:bar_top}).\\
Equally interesting is the structure around the top of the expected CdL bounce. In fig. \ref{fig:comb_scanplot}, there appears to be a very rapid movement towards $\sim 10^{-12}M_p$ (the value of the $\phi$ at the end of the false vacuum side of the solution). However, closer inspection in figure \ref{fig:regionB} shows that this variation is smooth, with the exception of a narrow region of overshoots among the mostly undershoot solutions. Since every overshoot/undershoot transition boundary implies the existence of a bounce, there are at least three bounces that start near $\phi(0) \sim 0.17M_{\rm{P}}$. The solutions must transition back to overshoots on reaching the barrier (since analytically solutions linearized about the barrier are known to be overshoots for $H_0 > H_{0\rm{crit}}$), thus a fourth solution necessarily exists. This solution is small amplitude and mostly confined to the top of the barrier. Thus, unlike for $H_0 < H_{0\rm{crit}}$, there are actually four non-trivial solutions in addition to the Hawking-Moss solution.
\\
Figure \ref{fig:sm_bounce_plots} shows example bounce solutions for $V_0 = 7.210\times 10^{-21}M_{\rm{P}}^4$ ($H_0 = 1.1937\times 10^{8}\rm{ GeV}$), which is just above the critical threshold. There are four solutions. Table \ref{tab:bounce_data} shows their end points and associated initial data compared to the relevant Hawking-Moss solution. For comparison, we show the same calculation assuming a fixed de Sitter background, neglecting the back-reaction terms, in table \ref{tab:bounce_data_fixed} Note that many of these initial conditions differ only in the sixth decimal place or more. To verify that this is not due to numerical error, the presence of an overshoot/undershoot boundary for each bounce solution was tested using the Fehlberg78 Runge-Kutta method\cite{Fehlberg78} at relative tolerance of $\varepsilon_{\rm{rel}} = 10^{-30}$ and absolute tolerance of $\varepsilon_{\rm{abs}} = 10^{-80}$ using arbitrary precision numbers with up to 100 decimal places of precision to reduce rounding error. Note that the extremely low absolute tolerance is only necessary near to the co-ordinate singularity at $\chi = 0$; for most of the solution's range the relative tolerance is far more important than the absolute tolerance for controlling numerical precision. Adaptive Runge-Kutta methods such as this vary the step size such that the error estimate $\Delta y$ for solution component $y$ satisfies:
\begin{equation}
|\Delta y| < |\varepsilon_{\rm{rel}}y + \varepsilon_{\rm{abs}}|
\end{equation}
at each step. This level of precision makes it easy to distinguish between solutions such as `CdL solution 3' and `CdL solution 4' in table \ref{tab:bounce_data} which start extremely close together, but terminate in very different places. The fact that the overshoot/undershoot boundaries persist when the relative tolerance is increased suggests that the effect is real, not a numerical artifact.
\begin{table*}
\begin{center}
\begin{tabular}{|l|l|l|l|l|}
\hline
\bf{Bounces with back-reaction} & $\phi(0)/\rm{GeV}$ & $\phi(\chi_{\rm{max}})/\rm{GeV}$ & \bf{Decay exponent}, $B$ & $H_0\chi_{\rm{max}}/\pi - 1$\\ \hline
CdL solution 1 &      $4.0589763\times 10^{17}$ & $2.5097992\times 10^6$     &$1808.261$ &$-3.99325\times 10^{-12}$ \\ \hline
CdL solution 2 &$5.5306295\times 10^{9}$   & $4.7385411\times 10^{9}$ &$12388.87$ &$-1.89647\times 10^{-19}$ \\ \hline
CdL solution 3 &$4.0588911\times 10^{17}$ & $4.7385591\times 10^{9}$ &$14197.13$ &$-3.99303\times 10^{-12}$ \\ \hline
CdL solution 4 &$4.0588976\times 10^{17}$ & $5.1096372\times 10^{9}$ &$14197.08$ &$-3.99303\times 10^{-12}$ \\ \hline
Hawking-Moss solution &$5.1096727\times 10^{9}$&$5.1096727\times 10^{9}$ &$12388.82$ &$-3.77098\times 10^{-19}$ \\
\hline
\end{tabular}
\end{center}
\caption{\label{tab:bounce_data}Table of initial and final values of the bounce solutions for $V_0 = 7.210\times 10^{-21} M_{\rm{P}}^{4}, (H_0 = 1.1937\times 10^{8} \rm{ GeV})$, together with the associated decay exponents. The ending values of $\chi_{\rm{max}}$ are all nearly the same as in the fixed background approximation, but this does not mean the effects of gravitational back-reaction are negligible. Note that CdL solution 2 and the Hawking-Moss solution have $\chi_{\rm{max}}$ significantly closer to the flat false vacuum result, as they probe only the barrier, which is closer to the false vacuum, while the other solutions probe the depth of the Standard Model potential and thus receive larger back-reaction corrections.}
\end{table*}
\begin{table*}
\begin{center}
\begin{tabular}{|l|l|l|l|l|}
\hline
\bf{Bounces with fixed dS background} & $\phi(0)/\rm{GeV}$ & $\phi(\chi_{\rm{max}})/\rm{GeV}$ & \bf{Decay exponent}, $B$\\ \hline
CdL solution 1 &      $6.5057883\times 10^{17}$ & $2.1207789\times 10^6$     &$1805.797$  \\ \hline
CdL solution 2 &$5.5306295\times 10^{9}$   & $4.7385412\times 10^{9}$ &$12388.88$  \\ \hline
CdL solution 3 &$6.5056176\times 10^{17}$ & $4.7385523\times 10^{9}$ &$14194.68$  \\ \hline
CdL solution 4 &$6.5056306\times 10^{17}$ & $5.1096372\times 10^{9}$ &$14194.62$ \\ \hline
Hawking-Moss solution &$5.1096727\times 10^{9}$&$5.1096506\times 10^{9}$ &$12388.82$  \\
\hline
\end{tabular}
\end{center}
\caption{\label{tab:bounce_data_fixed}Table of initial and final values of the bounce solutions using a fixed de Sitter background, for $V_0 = 7.210\times 10^{-21} M_{\rm{P}}^{4}, (H_0 = 1.1936\times 10^{8} \rm{ GeV})$, together with the associated decay exponents. As the metric is fixed at the de Sitter space of the false vacuum, $\chi_{\rm{max}} = \frac{\pi}{H_0}$ for all solutions.}
\end{table*}
Of course, the only solution of relevance for quantum tunneling is the solution with the smallest Euclidean action. In this case, the calculation appears to show that the largest amplitude solution (i.e. largest $\phi(0)$ on the true vacuum side and smallest $\phi(\chi_{\rm{max}})$ on the false vacuum side) has the lowest action. Thus the additional solutions, while interesting, are exponentially suppressed and do not contribute to the decay rate. However, this may not be the case for all values of $H_0$ or potential shapes.
\\
Note that for even larger values of $H_0$, even more solutions than these four begin to appear, which makes the process of finding and classifying them even more complicated, as it is difficult to guarantee that all solutions for a given $H_0$ have been found. The solutions found currently have been of higher action than the largest amplitude CdL solution, which most closely resembles the flat space bounce, but the existence of lower action solutions cannot be ruled out. If so, such solutions could potentially dominate vacuum decay at large Hubble rates. The existence of multiple solutions also raises the question of which solution, if any, approaches the $V_0 = 0$ bounce as $V_0\rightarrow 0$. The data in fig. \ref{fig:comb_scanplot} suggest that this would be the largest amplitude solution, since this most closely resembles the unique bounce found for $H_0 < H_{0\rm{crit}}$, and the other solutions appear to be related to narrow overshoot or undershoot regions. The behavior of this family of solutions in the $V_0\rightarrow 0$ limit is discussed in section \ref{sec:V0to0}.

\subsection{Vacuum decay rate as a function of $H_0$}
The existence of extra solutions for $H_0 > H_{0\rm{crit}}$ is a significant complication to the question of the vacuum decay rate for a given $H_0$. To make progress, we conjecture that the largest amplitude non-trivial solution, if it exists, will always have the smallest action, and thus dominate the decay rate. This seems plausible for two reasons: (1) the bounce of largest amplitude has the smallest $\phi$ value far outside its center, on the false-vacuum side. At first glance, it might be assumed therefore that Eq. \ref{eq:B_at_sol} predicts a \emph{higher} action since $V(\phi)$ is smaller in this exterior region. However, this naive conclusion neglects effects coming from $a^3(\chi)$. As it turns out, these effects more than cancel out the decrease in $V(\phi)$ and the effect of having $\phi(\chi)$ closer to the false vacuum is the decrease the action, in general. Reason (2) applies mainly to the small amplitude bounce: for $H_0 > H_{0\rm{crit}}$, the second eigenvalue for linear fluctuations about the Hawking-Moss solution is positive. This eigen-fluctuation corresponds to solutions which fluctuate about the top of the barrier, at a linearized level, with a shape similar to that of the non-linear solution which is close to the top of the barrier. Thus, it might be expected that this non-linear solution is a continuation in this `direction' of field configuration space, and has a larger action than the Hawking-Moss solution due to the positive eigenvalue. In order for non-linear solutions to have smaller action than the Hawking-Moss, we would expect to encounter a stationary point in between, and since this bounce is presumably the closest stationary point in this 'direction', it would be expected to have larger action.\\
Both these arguments are admittedly rather hand-waving - the first assumes that the competition between changes in $V(\phi)$ and $a^3(\chi)$ is always `won' by $a^3(\chi)$ in such a way as to decrease the action for larger amplitude bounces. The second relies on arguing that the action of non-linear solutions about the top of the barrier should behave qualitatively similar to that of linearized solutions, which is by no means certain without knowing the full structure of stationary points (including, possibly, non O(4) symmetric stationary points).\\
However, assuming this conjecture holds, it is relatively easy to extract only the largest amplitude solution for a given $H_0$. The point at which the Hawking-Moss solution begins to dominate is then the value of $H_0$ at which the Hawking-Moss solution and the largest amplitude CdL solution have equal action. We subsequently denote this value as $H_{0\rm{cross}}$. The total action is plotted in figure \ref{fig:action_plot}, giving $H_{0\rm{cross}} = 1.931\times 10^8 \rm{ GeV} (V_0 = 1.887\times 10^{-20}M_{\rm{P}}^4)$. One immediate observation is how flat the action curve appears. In fact, it does slope very slowly (see figure \ref{fig:action_plot}). In light of this, it is a reasonably good approximation to say that the decay rate exponent is the same as the $V_0 = 0$ case for $H_0 < H_{0\rm{cross}}$, and the same as the Hawking-Moss decay exponent for $H_0 > H_{0\rm{cross}}$. As can be seen from figure \ref{fig:action_plot}, there is no special behavior at $H_{0\rm{crit}}$, unlike in the polynomial case considered.

\section{Flat false vacuum limit\label{sec:V0to0}}
As mentioned earlier in the paper, there is a significant problem in the $V_0\rightarrow 0$ limit, in that the decay exponent arises due to a cancellation of two large numbers:
\begin{equation}
B = \frac{24\pi^2 M_{\rm{P}}^4}{V_0} - 2\pi^2\int_{0}^{\chi_{\rm{max}}}\dd\chi a^3(\chi)V(\phi(\chi)).
\end{equation}
The first term diverges as $V_0\rightarrow 0$, but the result is known to be finite there, thus the second term must also diverge in such a way that the overall result is finite. This means that computing each part separately is highly inaccurate, as a result of round off error. A similar problem occurs in the limit where $V_0 \gg |\Delta V(\phi_{\rm{HM}})|$, for which Eq. (\ref{eq:B_HM}) becomes inaccurate, a problem which can be cured by using a Taylor series approximation.\\
The way this problem is solved has already been discussed in section \ref{sec:numerical}. Here we attempt to provide an answer to a reasonable question - what happens to bounce solutions in the $V_0\rightarrow 0$ limit? Does the decay exponent, $B$, smoothly approach the $V_0 = 0$ value? For both the polynomial potential and the standard model potential considered in this paper, the answer appears to be yes, at least at the numerical level. We would like to put this question of somewhat firmer grounds analytically, however.\\
To do this, we analyze the conditions under which $B_1$, $B_2$, and $B_3$, discussed in section \ref{sec:numerical} converge to the $V_0 = 0$ result. We first need to state precisely what this means: naively, we would say that the CdL bounce should approach the $V_0 = 0$ bounce as $V_0\rightarrow 0$. However, since there are potentially many CdL bounces, and bounces for different $V_0$ are not defined on the same manifold, it is not immediately obvious if there is any meaning to talk about the `same' bounce at different values of $V_0$. For the purposes of this paper, we will merely assume there exists a sequence of bounce solutions with different $V_0$ whose limit as $V_0\rightarrow 0$ is the $V_0 = 0$ bounce. The question is then whether the action of this sequence approaches the $V_0 = 0$ action if the solutions approach the $V_0 = 0$ bounce.\\
\subsection{Vanishing of $B_3$\label{sec:B3vanish}}
The first thing to note here is that provided $\chi_{\rm{max}}\rightarrow \infty$, $\phi_{H_0}(\chi) \rightarrow \phi_{0}(\chi)$, $a_{H_0}(\chi)\rightarrow a_0(\chi)$ smoothly (where $\phi_{H_0}(\chi),a_{H_0}(\chi)$ describe the bounce solution for a given $H_0$ and $\phi_0(\chi), a_0(\chi)$ describe the $V_0 = 0$ solution), then $B_1\rightarrow B_0$, the flat-false-vacuum decay exponent. Thus, for the limit to be smooth, both $B_2$ and $B_3$ must vanish in the $V_0\rightarrow 0$ limit. The condition for $B_3$ to vanish is the simplest to establish, so we will consider that first:
\begin{align}
B_3 =& \frac{24\pi^2M_{\rm{P}}^4}{V_0}\nonumber\\
& - 2\pi^2\int_{0}^{\chi_{\rm{max}}(H_0)} \left(\frac{1}{H_0}\sin(H_0(\chi_{\rm{max}} - \chi))\right)^3V_0\nonumber\\
 =& 2\pi^2\int_{0}^{\frac{\pi}{H_0}}\dd\chi\left(\frac{1}{H_0}\sin(H_0\chi)\right)^3V_0\nonumber\\
& - 2\pi^2\int_{0}^{\chi_{\rm{max}}(H_0)}\dd\tilde{\chi} \left(\frac{1}{H_0}\sin(H_0\tilde{\chi})\right)^3V_0,
\end{align}
which is a `near cancellation' of the divergent false vacuum action. Note this equation alone is not guaranteed to give a finite answer - it depends on the $\chi_{\rm{max}}(H_0)$ function (implicit in writing down this function is the assumption that the aforementioned sequence of bounces vary smoothly). For $B_3$ to vanish, the condition is that $1  + \cos(H_0\chi_{\rm{max}}(H_0))\rightarrow 0 $ faster than $H_0$, which is equivalent to saying:
\begin{equation}
\lim_{H_0\rightarrow 0} H_0\chi_{\rm{max}}(H_0) = (2n + 1)\pi, n = 0,1,2,\ldots\label{eq:chimax_limit}
\end{equation}
Note that since $\phi_0(\chi)\rightarrow 0$ as $\chi\rightarrow \infty$ for the $H_0 = 0$ solution, and $\phi_{H_0}(\chi)$ is assumed to smoothly approach $\phi_0(\chi)$, the majority of the $[0,\chi_{\rm{max}}]$ domain will have negligible $V(\phi(\chi))$, and thus one expects $\chi_{\rm{max}}$ to be similar to $\frac{\pi}{H_0}$, not $\frac{3\pi}{H_0}$ or some other odd integer multiple (this could only happen if there were significant back-reaction over the majority of the domain, which is not the case if the solution is smoothly approaching $\phi_0(\chi)$, which is approximately zero over the majority of its (infinite) domain). Consequently, we expect to be able to write down a power series:
\begin{equation}
\chi_{\rm{max}}(H_0) = \frac{\pi}{H_0} + \alpha_0 + \alpha_1 H_0 + O(H_0^2),\label{eq:chimax_series}
\end{equation}
where $\alpha_i$ are some undetermined constants, at least in some neighborhood of $H_0 = 0$. In fact, it is possible to determine $\alpha_0$ from the flat-false vacuum solution alone. In that limit, one finds the scale factor, $a_0(\chi)$, to be:
\begin{align}
a_{0}(\chi) &= \lim_{H_0\rightarrow 0}\left(\frac{1}{H_0}\sin(H_0(\chi_{\rm{max}}(H_0) - \chi))\right) + \delta a_0(\chi)\nonumber\\
&= \chi - \alpha_0 + \delta a_0(\chi).
\end{align}
Note that because of the way it is defined, $\delta a_0(\chi) \rightarrow 0$ as $\chi\rightarrow \infty$ (this wouldn't have been the case if the $\delta a_{H_0}(\chi)$ splitting had been done differently, which is why Eq. \ref{eq:asplit} is the most `natural' choice). This means:
\begin{equation}
\alpha_0 = \lim_{\chi\rightarrow\infty}(\chi - a_0(\chi)).
\end{equation}
This is in fact a fairly stable numerical calculation to do. For the values of the top quark and Higgs mass used in this paper, we find $\alpha_0 = -0.2559M_{P}^{-1}$. The fact that this is negative is expected - it is a consequence of the fact that the bounce solution has negative curvature in the interior of the bounce due to the potential being negative there. This results in a (very) brief period of exponential growth of $a_0(\chi)$ in the interior of the bounce, meaning that at large $\chi$, $a_0(\chi)$ is always slightly larger than $\chi$ and goes as $a_0(\chi) \sim \chi - \alpha_0$. This translates into a negative $\alpha_0$, which can be interpreted physically as characterizing the `net' back-reaction of the bounce.\\
Of course, this only gives the condition for $B_3$ to be finite in the $H_0\rightarrow 0$ limit. It does not prove that the requisite one-parameter family of solutions exists. However, if it does, then the condition $\lim_{H_0\rightarrow 0}H_0\chi_{\rm{max}}(H_0) = \pi$ must be satisfied, otherwise the action of the family of bounces diverges.\\
To ascertain whether we expect this to be the case, consider changing to a co-ordinate system $x = H_0\chi$. In this co-ordinate system, as $\phi_{H_0}(x)\rightarrow \phi_0(x)$, it becomes an infinitely narrow spike because $\phi_0(\chi)$ approaches a fixed `width' (e.g, $\chi$ at which the bounces reaches half its maximum value), implying that the width in the $x$ co-ordinates of $\phi_0(x)$ decreases with decreasing $H_0$ and is proportional to $H_0$. The scale factor equation in this co-ordinate system becomes:
\begin{equation}
a'' = -\frac{a}{3M_{\rm{P}}^2}\left(\phi'^2 + 3M_{\rm{P}}^2 + \frac{\Delta V(\phi)}{H_0^2}\right).
\end{equation}
For a `narrow spike' solution, $\phi'^2 = \Delta V(\phi) = 0$ over most of the range of integration, increasingly so in the $H_0\rightarrow 0$ limit. Thus, far outside the bounce, the solution satisfies $a'' = -a$ with solution $a(x) = \sin(x+\varphi)/H_0$. The phase $\varphi$ is fixed by asymptotic matching to the interior solution. Since the region in which the interior solution is not negligible shrinks to zero in the $H_0\rightarrow 0$ limit, then so does $\varphi$ and the domain size approaches $\pi$, implying $\lim_{H_0\rightarrow 0}H_0\chi_{\rm{max}}(H_0) = \pi$.\\
\begin{figure}
\includegraphics[width=0.5\textwidth]{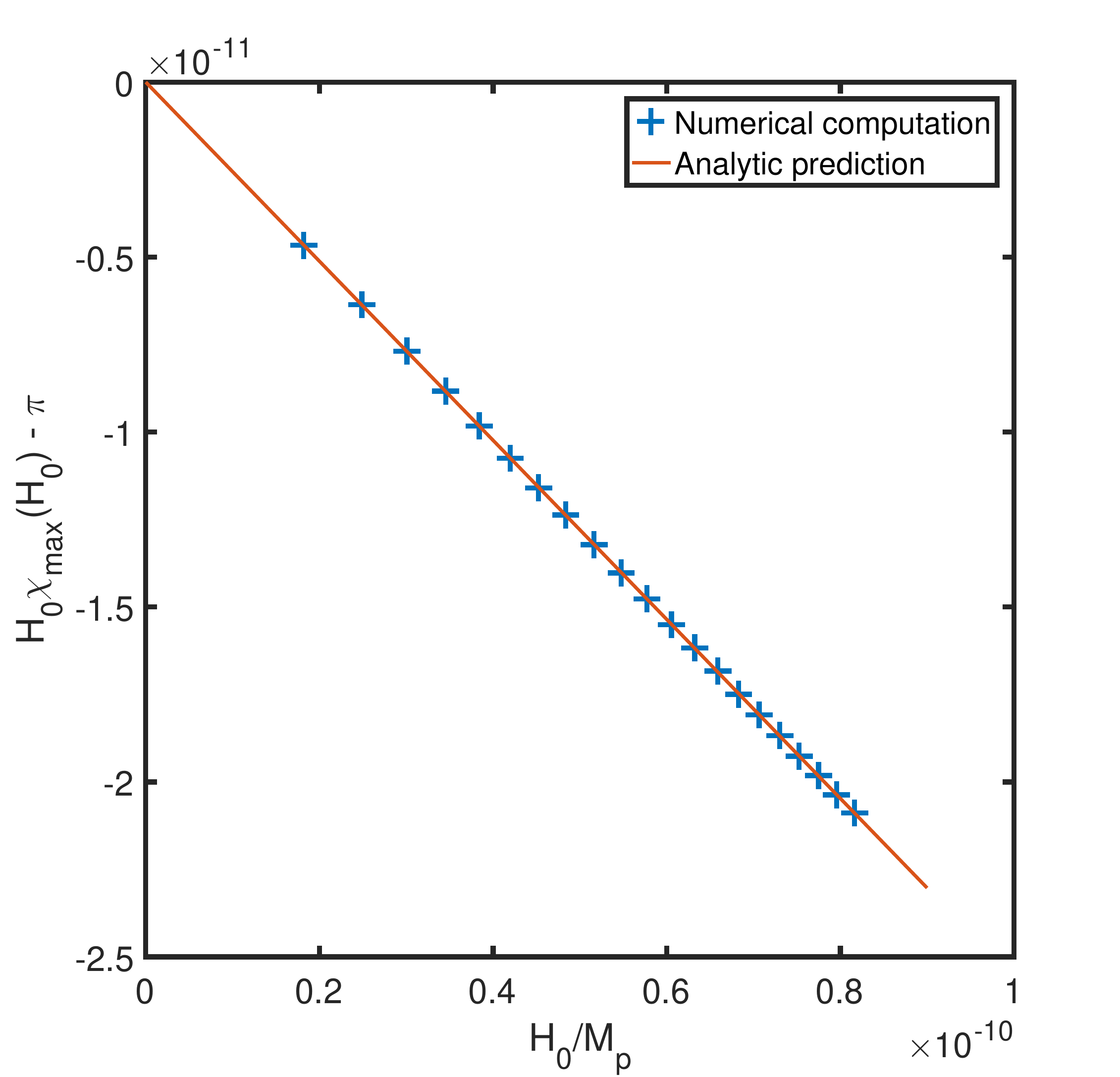}
\caption{\label{fig:chimax_plot}Plot of $H_0\chi_{\rm{max}}(H_0) - \pi$ against $H_0$ for a range of CdL bounces in the Standard Model, computed numerically (crosses) and compared to the analytic predicion $\chi_{\rm{max}}\approx \frac{\pi}{H_0} - 0.2559 M_{P}^{-1} + \ldots$ of Eq. (\ref{eq:chimax_series}).}
\end{figure}
The question then becomes whether such a family exists. We have not been able to provide a satisfactory formal proof of this, however, at least in the potentials we have considered, it appears to be plausible. Figure \ref{fig:chimax_plot} shows $H_0\chi_{\rm{max}}(H_0) - \pi$ for a family of CdL solutions in the Standard Model effective potential, plotted against the analytic prediction for $\chi_{\rm{max}}(H_0)$ in  Eq. (\ref{eq:chimax_limit}). The slope and y-intercept of the resulting line are consistent with satisfying the $H_0\chi_{\rm{max}}(H_0) - \pi \rightarrow 0$ relationship, to within the error of the coefficients of a linear model. The least squares regression fit $H_0\chi_{\rm{max}}(H_0) - \pi = a + bH_0$ gives, for these data, $a = (-0.1 \pm 1.4)\times 10^{-16}$ and $b = -0.255894\pm 0.0000025 M_{\rm{P}}^{-1}$. This gives a limit for $H_0\chi_{\rm{max}}(H_0) - \pi$ consistent with $0$ and a slope consistent with $\alpha_0 = -0.2559$, as extracted from the $H_0 = 0$ solution. On this basis, we regard it as extremely plausible (though still unproven) that there exists a family of bounce solutions satisfying the requisite condition on $\chi_{\rm{max}}$, at least in the Standard Model potential.

\section{Discussion\label{sec:discussion}}
The most important observation is that the Standard Model is not a typical potential: there is no continuous transition between CdL instantons and Hawking-Moss instantons as occurs in some model potentials. This may call into question arguments based on direct analogy between the Standard Model and simple models, as it is clear the behavior of tunneling in the Standard Model is strongly dependent on the shape of the potential. However, our results suggest that the dependence of CdL bounces on the Hubble rate is extremely weak for Hubble rates a long way below the scale of the CdL bounce itself. From an intuitive point of view this makes sense - the Standard Model bubble nucleation process occurs on scales only an order of magnitude smaller than the Planck scale\cite{PhysRevD.77.025034,PhysRevD.95.025008}, and so barely sees the curvature associated to the Hubble rate unless it is also close to that scale. Since the Planck results imply a tensor-to-scalar ratio $r < 0.11$\cite{planck}
, corresponding to $H_0 < 7.9\times 10^{13}\rm{ GeV}$, it should be a good approximation for most inflationary models to use the $V_0 = 0$ bounce for Hubble rates before the crossing point, $H_0 < H_{0\rm{cross}} = 1.931\times 10^{8} \rm{ GeV}$
, and the Hawking-Moss (or Fokker-Planck) analysis above this threshold. This assumes, of course, that none of the extra solutions which appear have lower action than the largest amplitude CdL or Hawking-Moss solution. Neglecting the weak $H_0$ dependence and assuming a constant $B(H_0) = B_{\rm{fv}}$, the approximate location of the cross-over is:
\begin{equation}
V_{0\rm{cross}} \approx \frac{1}{2}\Delta V(\phi_{\rm{HM}})\left(-1 + \sqrt{1 + \frac{96\pi^2 M_{\rm{P}}^4}{\Delta V(\phi_{\rm{HM}})B_0}}\right),
\end{equation}
where $B_0$ is the $V_0 = 0 $ decay exponent.\\
The emergence of extra solutions in the Standard Model potential immediately complicates calculations of the vacuum decay rate. Ostensibly, the bounce with lowest action should always dominate. However, this may not be straightforward to identify, as there appears to be no obvious way of predicting how many CdL solutions are expected for a given Hubble rate. As the Hubble rate was raised in our analysis, we found that more and more solutions appeared. Although for $H_0 < H_{0\rm{crit}}$ only a single CdL bounce and the Hawking-Moss instanton were found, the existence of narrow regions of overshoots among regions which otherwise look like undershoots, or vice versa (fig. \ref{fig:regionB} is a typical example) for $H_0 > H_{0\rm{crit}}$ means that we cannot rule out the existence of other, unknown, solutions for \emph{any} $H_0$. We hypothesize that these extra solutions only appear for $H_0 > H_{0\rm{crit}}$, but can offer no proof of this, and it could well turn out to be false.
\\
In all the cases we studied, the action of the extra solutions was found to be larger than the largest amplitude CdL solution (which is the solution we hypothesize to be in the family that continuously deforms into the $V_0 = 0$ solution, with other solutions disappearing as $H_0$ is lowered). We hypothesize, therefore, that this largest amplitude solution always has the lowest action, and thus the other solutions should be irrelevant to vacuum decay. There are other reasons to think that this is the case: the existence and form of these solutions probably depends strongly on the shape of the potential. However, the precise shape of the potential is not a gauge invariant property; only physical observables such as the locations of its stationary points are\cite{NIELSEN1975173}. It may be the case, therefore, that the existence and nature of these solutions depends on the choice of gauge: in which case they may not be physical and we would not expect them to change the decay rate be possessing a lower action, which \emph{should} be a guage invariant quantity.
\\
This is conjecture, however, and doesn't provide an explanation for why the lower amplitude solutions should have higher action. Naively, the smaller amplitude looks like it should actually decrease the decay exponent because $V(\phi)$ stays larger in the exterior, which should decrease the decay exponent, according to Eq. (\ref{eq:decayExponent}). Indeed, this \emph{does} reduce $B_1$ (Eq. (\ref{eq:B1}) ) for the solutions we considered, but this was more than compensated for by an increase in $B_2$ (intuitively, the change in the geometry is more significant), which is actually significant for the `extra' solutions, unlike the case in the large amplitude solutions, because the asymptotic geometry is closer to the Hawking-Moss solution than it is to the false vacuum solution. However, there remains no proof at this point that the smaller-amplitude solutions do always have higher action. The emergence of new solutions as $H_0$ is raised means that the appearance of extra solutions with lower action, which would consequently dominate vacuum decay, cannot be ruled out. If a narrow range of initial values, $\phi_0$, can produce overshoots or undershoots while everything around it can produce undershoots, then there is no obvious way of knowing that all CdL solutions have been found.
\\
Our approach was to search for discontinuities in the scan plots like fig. \ref{fig:comb_scanplot}, and examine values of $\phi_0$ around them until the narrow regions were found. Not all discontinuities are so obvious, however, such as that in fig. \ref{fig:regionB}. For there, it was necessary to `follow' the solutions we found along to their termination value, $\phi(\chi_{\rm{max}})$ and search at higher resolution around that point to uncover the narrow regions where additional solutions lay. This process was repeated until all overshoot-undershoot transitions could be matched with a transition on the other side of the barrier. However, this procedure does not guarantee that additional solutions, perhaps `disconnected' from the ones already found, do not exist.
\\
For comparison, we computed bounce solutions for the Standard Model both with full back-reaction, and with the `fixed background' approximation, that is, assuming that the metric is unchanged from that of the de Sitter background in the false vacuum. The use of a fixed de Sitter background did not change the conclusion that extra solutions appear, and their actions are only slightly different (see table \ref{tab:bounce_data_fixed}). The difference in decay exponents between the fixed de Sitter and full back-reaction cases is essentially the same as is found in the flat false vacuum ($V_0 = 0$) case between the cases when back-reaction is included and neglected\cite{PhysRevD.95.025008}, which is expected as the decay exponents for the largest amplitude (`CdL solution 1') solutions do not deviate significantly from the $V_0 = 0$ value. CdL solutions 3 and 4 display similar behavior. The most significant difference between the two cases (fixed vs non-fixed background) is that the bounce solutions in the fixed case have a higher peak $\phi(0)$. This behavior was also seen in the $V_0 = 0$ case and is caused by gravitational back-reaction altering the bounce scale that dominates the decay rate\cite{PhysRevD.77.025034,PhysRevD.95.025008,Salvio2016}.
\\
As mentioned earlier, Joti et al.\cite{Joti2017} recently published a paper discussing vacuum instability in the Standard Model during inflation, using a perturbative method. They consider vacuum instability in the Standard Model, in the regimes $H_0\ll H_{0\rm{crit}}, H_0\sim H_{0\rm{crit}}$ and $H_0 \gg H_{0\rm{crit}}$ respectively. They argue that by Taylor expanding around the top of the barrier for $H_0\sim H_{0\rm{crit}}$, one can find bounces satisfying:
\begin{align}
\Delta =& -\frac{(\phi(0) - \phi_{\rm{HM}})^2}{14H_0^2}\left[V^{(4)}(\phi_{\rm{HM}}) + \frac{V^{(3)}(\phi_{\rm{HM}})^2}{12H_0^2}\right]\label{eq:delta}\\
S =& S_{\rm{HM}} + \frac{2\pi^2(\phi(0) - \phi_{\rm{HM}})^2\Delta}{15H_0^2}\\
\Delta \equiv & 4 + \frac{V''(\phi_{\rm{HM}})}{H_0^2}.
\end{align}
One can re-write $\Delta$ as $\Delta = \frac{4(H_0^2 - H_{0\rm{crit}}^2)}{H_0^2}$. Joti et al. find that the RHS of Eq. (\ref{eq:delta}) is positive in the Standard Model, which implies that CdL bounces exist for $H_0 > H_{0\rm{crit}}$, with action larger than the Hawking-Moss action. On this basis, they argue that for $H_0 > H_{0\rm{crit}}$, Hawking-Moss solutions should dominate vacuum decay, with actions in the vicinity of $B \sim 13000$ for $H_0 = H_{0\rm{crit}}$.
\\
Our results agree with the conclusion that CdL bounces exist for $H_0 > H_{0\rm{crit}}$. In fact, we would identify these solutions with CdL solution 2 (see tables \ref{tab:bounce_data} and \ref{tab:bounce_data_fixed}), which we found numerically. However, our numerical search demonstrated the existence of \emph{multiple} CdL bounces, in particular, there is a large amplitude CdL bounce closely matching the $V_0 = 0$ case in its interior, but differing in the exterior (this is called CdL solution 1 in tables tables \ref{tab:bounce_data} and \ref{tab:bounce_data_fixed}), which has a \emph{smaller} action than the Hawking-Moss solution at $H_0 = H_{0\rm{crit}}$. This solution, with $B \approx 1800$, should dominate vacuum decay around $H_0 \sim H_{0\rm{crit}}$ and slightly above, not the Hawking-Moss solution and the extra bounce solutions. This means that Hawking-Moss solutions do not dominate until $H_0 = H_{0\rm{cross}}$, when $S_{\rm{HM}} \sim 1800$. In the case we considered, this occurs at $H_{0\rm{cross}} = 1.9313\times 10^{8} \rm{ GeV}$, higher than, but the same order of magnitude as $H_{0\rm{crit}} = 1.1931\times 10^8 \rm{ GeV}$. In practice, however, the effect of this is simply to shift the threshold at which Hawking-Moss domination begins. For small $H_0 \ll H_{0\rm{crit}}$, and $H_0 > H_{0\rm{cross}}$, our results agree with the conclusions of \cite{Joti2017}.

\section{Conclusion\label{sec:conc}}
The most practical conclusion that can be drawn from these results is that the largest amplitude CdL solution has almost identical action to the $V_0 = 0$ bounce, and this action depends only very weakly on the Hubble rate. Thus, below the cross-over threshold, it is a good approximation to simply use the $V_0 = 0$ bounce action. This means that the question of the vacuum decay rate becomes one of comparison between the largest amplitude CdL (which we conjecture to have the lowest action of the CdL solutions) and the Hawking-Moss solution. There is no smooth transition between them - whichever has the lowest action will dominate.\\
Importantly, however, we have found that the Standard Model effective potential leads to a very rich set of bounce solutions contributing to vacuum decay. It is likely that the form and nature of these solutions depends strongly on the shape of the potential. Since the effective potential of the Standard Model couplings will have curvature-corrections in de-Sitter space, further work is needed to establish the form of these solutions - if they exist - in a potential with appropriate curvature dependent energy scale. The effect of non-minimal coupling, as was considered in Ref. \cite{Joti2017} for example, would be of particular interest. Despite this, our results demonstrate vacuum instability in the Standard Model is not straightforward when gravity is included. Not only are the results quantitatively and qualitatively different to the flat false vacuum case, but there are new solutions which are not present in the $V_0 = 0$ case.\\
Additionally, we showed that in the $V_0\rightarrow 0$ limit, the decay exponent will smoothly approach the $V_0 = 0$ exponent, provided a family of bounce solutions exist which approach the $V_0 = 0$ bounce, and that $\chi_{\rm{max}}$ for this family approaches zero in the manner of Eq. (\ref{eq:chimax_limit}). Although we have not proven that such a family always exists, we have shown that its existence is plausible, and we have numerically located a family of bounce solutions for which $\chi_{\rm{max}}$ appears to satisfy Eq. (\ref{eq:chimax_series}) to a high degree of accuracy. This suggests that the $V_0 \rightarrow 0$ limit is smooth in the Standard Model, which justifies neglecting the small cosmological constant observed today in calculations of the vacuum decay rate.\\
\\
Our results broadly agree with those of \cite{Joti2017} regarding the numerical value for the decay rate in the Standard Model. However, we found additional CdL bounce solutions that are not present in simpler polynomial models, which change the behavior of the CdL solution as a function of the background Hubble rate, $H_0$, particularly in the vicinity of the critical value, $H_{0\rm{crit}}$. In particular, the largest amplitude CdL solution does not merge with the Hawking-Moss solution at $H_{0\rm{crit}}$, and instead persists at higher values of the Hubble rate.\\
\\
These results also highlight that the critical threshold, usually fixed by the eigenvalues of the Hawking-Moss solution, is extremely important in some potentials, and virtually irrelevant in others (such as the Standard Model). The precise criteria that govern the existence of CdL solutions, including `extra' solutions not present when $V_0 = 0$, remain unclear. Jensen and Steinhardt pointed out the importance of the $H_0 < H_{0\rm{crit}}$ condition in reference \cite{Jensen1984176}. However, this does not result in a straightforward bound: $H_0 < H_{0\rm{crit}}$ guarantees a CdL bounce exists, but $H_0 > H_{0\rm{crit}}$ does not rule it out\cite{PhysRevD.69.063518}. It has been suggested that some sort of average over the barrier should be the criteria instead\cite{PhysRevD.71.044014,PhysRevD.69.063518}. Since bounces violating the $H_0 < H_{0\rm{crit}}$ bound are present in the Standard Model, and not just model potentials, a greater understanding of how and why CdL solutions exist for a given potential is necessary.
\subsection{Acknowledgments}
We would like to thank Alberto Salvio for useful comments. AR was supported by STFC grant ST/L00044X/1 and SS by the President's PhD Scholarship.

\appendix
\section{Vanishing of $B_2$\label{sec:B2vanish}}
The $B_2$ contribution to the action can be shown to vanish in the $H_0$ limit, but the manner in which it vanishes depends on the second derivative of the potential in the false vacuum. Taking the limit as $H_0\rightarrow 0$, $B_2$ reduces to:
\begin{align}
B_2 =& -6\pi^2M_{\rm{P}}^2\lim_{H_0\rightarrow 0}H_0^2\int_{0}^{\frac{\pi}{H_0} + \alpha_0 + \ldots}\dd\tilde{\chi}\left[3(\chi - \alpha_0)^2\delta a_{0}(\chi)\right.\nonumber\\
& \left.+ 3(\chi - \alpha_0)\delta a_{0}^2(\chi) + \delta a_0^3(\chi) + O(H_0^2)\right].
\end{align}
This will be zero if the integral part is finite. To establish this, we need to know the asymptotic behavior of $\delta a_0(\chi)$. As $\chi\rightarrow \infty$, the $V_0 = 0$ solution approaches the false vacuum, so the potential is well approximated by $V(\phi) = \frac{1}{2}m^2(\phi - \phi_{\rm{fv}})^2 + \ldots$ (if $m^2 = V''(\phi_{fv}) > 0$ - we will consider the $V''(\phi_{\rm{fv}}) = 0$ case separately). The scalar field equation becomes, approximately:
\begin{equation}
\Delta\ddot{\phi} + \frac{3}{\chi}\Delta\dot{\phi} - m^2\Delta\phi = 0,
\end{equation}
where $\Delta\phi = \phi - \phi_{\rm{fv}}$ (here we neglect quadratic terms like $\delta a_0(\chi) \Delta \phi$). This has a known solution in terms of Bessel functions. In particular, to match the boundary conditions, it must be of the form of the modified Bessel function of the second kind, $\Delta\phi\propto K_{1}(m\chi)/\chi$. Asymptotically, this is exponential decay:
\begin{equation}
\phi(\chi) = \frac{C}{\chi^{3/2}}\exp(-m\chi).
\end{equation}
$\delta a_0(\chi)$ then satisfies:
\begin{align}
\delta \ddot{a}_0 =& - \frac{(\chi + \delta a_0)}{3M_{\rm{P}}^2}\left(\frac{C^2m^2}{\chi^3}e^{-2m\chi} + \frac{1}{2}m^2 \frac{C^2}{\chi^3}e^{-2m\chi}\right.\\
&\left.+ O\left(\frac{e^{-2m\chi}}{\chi^{4}}\right)\right)
\end{align}
which means that it approaches a constant (0, because of the boundary condition on $\delta a_0(\chi)$) exponentially fast. This is sufficient to ensure that the integral is finite, so $B_2\rightarrow 0$.\\
The case of $m^2 = 0$ is slightly more complicated, since the equation is not linearizable in that case. However, it will approximately satisfy:
\begin{equation}
\Delta\ddot{\phi} + \frac{3}{\chi}\Delta\dot{\phi} = 0,
\end{equation}
for large $\chi$, which gives $\phi = \frac{C}{\chi^2}$. Using the same argument as before, this means:
\begin{align}
\delta \ddot{a}_0 \approx& \frac{\chi}{3M_{\rm{P}}^2}\left(\frac{4C^2}{\chi^6} + \frac{V^{(4)}(\phi_{\rm{fv}})}{24} \frac{C^4}{\chi^8}\right)\\
\implies & \delta a_0(\chi) \approx  \frac{C^2}{9M_{\rm{P}}^2\chi^3}.
\end{align}
Note that a $\frac{1}{\chi^3}$ dependence for $\delta a_0(\chi)$ in this case implies that the integral logarithmically diverges, as $(\chi - \alpha_0)^2\delta a_0(\chi) \sim \frac{1}{\chi}$. However, since the upper limit is asymptotically $\frac{\pi}{H_0}$, $B_2$ still approaches zero, asymptotically as $B_2\sim \lim_{H_0\rightarrow 0} H_0^2\log(H_0) \rightarrow 0$. Thus, $B_2$ always vanishes in the $H_0\rightarrow 0$ limit, but does so more slowly if $m^2 = 0$, approaching zero as $\sim H_0^2\log(H_0)$ rather than $\sim H_0^2$.

\section{False vacuum overshoots\label{sec:fvovershoots}}
In the paper it was claimed that solutions sufficiently close to the true vacuum are always overshoots. We now prove this assertion. Consider linearizing about the false vacuum. The scale factor is approximately $a = \frac{1}{H_0}\sin(H_0\chi)$ since we remain infinitesimally close to the false vacuum, and the scalar field satisfies:
\begin{equation}
\Delta\ddot{\phi} + 3H_0\cot(H_0\chi)\Delta\dot{\phi} - V''(\phi_{\rm{fv}})\Delta\phi = 0.
\end{equation}
We considered this equation earlier in the paper (see Eq. (\ref{eq:deltaphi}) ). Transcribing the Hypergeometric function solution we find:
\begin{align}
\Delta\phi(\chi) =& \Delta\phi_0\text{ }{}_2F_1\left(\frac{3}{2} + \alpha,\frac{3}{2} - \alpha,2,\sin^2\left(\frac{H_{0}\chi}{2}\right)\right).\\
\sim & -\frac{4\Delta\phi_0 \cos\left(\pi\sqrt{\frac{9}{4} - \frac{V''(\phi_{\rm{fv}})}{H_{0}^2}}\right)}{(2 - \frac{V''(\phi_{\rm{fv}})}{H_{0}^2})\pi(\pi - H_{0}\chi)^2}.\label{eq:fvdiverge}\\
\alpha =& \sqrt{\frac{9}{4} - \frac{V''(\phi_{\rm{fv}})}{H_{0}^2}}.\nonumber
\end{align}
To establish that this always overshoots, it is necessary to show that (1) Eq. (\ref{eq:fvdiverge}) diverges with the same sign as $\Delta\phi_0$ (that is, $\Delta\phi(\chi)/\Delta\phi_0$ diverges with positive coefficient) and (2) there are no turning points in the solution (as these could imply a solution which oscillates before diverging, which would provide a bound on $\phi_0$ for an oscillating bounce solution instead of the CdL bounce we are interested in).\\
The relevant hypergeometric differential equation here can be written as:
\begin{equation}
z(1-z)\frac{\dd^2\Delta\phi}{\dd z^2} + (2-4z)\frac{\dd\Delta\phi}{\dd z} - \left(\frac{9}{4} - \alpha^2\right)\Delta\phi = 0,\label{eq:hypergeom}
\end{equation}
where $z = \sin^2\left(\frac{H_0\chi}{2}\right)$. We aim to show that this solution diverges at $z = 1$ without encountering any turning points. Note that the following argument only works if $V''(\phi_{\rm{fv}}) > 0$ - the $V''(\phi_{\rm{fv}}) = 0$ case cannot be linearized. Assuming $V''(\phi_{\rm{fv}}) > 0$, we always have $\alpha^2 < \frac{9}{4}$. So, at a turning point of the solution:
\begin{equation}
\frac{\dd^2\Delta\phi}{\dd z^2} = \frac{\left(\frac{9}{4}-\alpha^2\right)}{z(1-z)}\Delta\phi.\label{eq:increasing}
\end{equation}
Thus, for $\alpha^2 < \frac{9}{4}$, turning points always have the same sign second derivative as the sign of $\Delta\phi$. In particular, if $\Delta\phi$ is positive, they will always be minima. Consider $\Delta\phi_0 > 0$: $\Delta \phi$ starts positive and initially increases (by Eq. (\ref{eq:phiboundary}), since $V'(\phi_{\rm{fv}} + \Delta\phi_0) > 0$ for sufficiently small positive $\Delta\phi_0$). This means it cannot ever encounter a minimum other than the initial minimum at $z = 0$, since that would require first encountering a maximum for positive $\Delta\phi$, which is impossible by Eq. (\ref{eq:increasing}). The same applies in reverse for $\Delta\phi_0 < 0$: there is an initial maximum and the solution can never encounter a minimum, so cannot encounter a second maximum either. Consequently, $z = 0$ is the only stationary point and the solution monotonically increases/decreases according to the sign of $\Delta\phi_0$. It must also therefore diverge. In fact, the only regular solutions to Eq. (\ref{eq:hypergeom}) occur at specific (eigen-)values of $\alpha^2$ such that the coefficient of the divergent part of Eq. (\ref{eq:divergence}) is zero:
\begin{equation}
\frac{\Delta\phi(z)}{\Delta\phi_0} \sim \frac{\cos(\pi\alpha)}{\left(\frac{1}{4}-\alpha^2\right)\pi(1-z)},\label{eq:divergence}
\end{equation}
near $z = 1$. $\Delta\phi/\Delta\phi_0$ is positive near $z = 1$, since it is positive and increasing near $z = 0$ and cannot have a stationary point except at $z=0$. This can also be shown explicitly: for $\alpha^2 < 0$, $\cos(\alpha\pi) = \cosh(|\alpha|\pi) > 0$; for $0 < \alpha^2 < \frac{1}{4}$, $\cos(\pi\alpha) > 0$ and $\frac{1}{4} - \alpha^2 > 0$; for $\frac{1}{4} < \alpha^2 < \frac{9}{4}$, $\cos(\pi\alpha) < 0$ but $\frac{1}{4} - \alpha^2 < 0$ too; for $\alpha^2 = \frac{1}{4}$, $\lim_{\alpha\rightarrow \frac{1}{2}}\frac{\Delta\phi(z = 1)}{\Delta\phi_0} = 1$. Hence the solution always diverges on the same side as $\Delta\phi_0$, and never has any turning points. It is, therefore, an overshoot.

\FloatBarrier
\bibliography{smvdidSs_references}

\begin{thebibliography}{41}%
\makeatletter
\providecommand \@ifxundefined [1]{%
 \@ifx{#1\undefined}
}%
\providecommand \@ifnum [1]{%
 \ifnum #1\expandafter \@firstoftwo
 \else \expandafter \@secondoftwo
 \fi
}%
\providecommand \@ifx [1]{%
 \ifx #1\expandafter \@firstoftwo
 \else \expandafter \@secondoftwo
 \fi
}%
\providecommand \natexlab [1]{#1}%
\providecommand \enquote  [1]{``#1''}%
\providecommand \bibnamefont  [1]{#1}%
\providecommand \bibfnamefont [1]{#1}%
\providecommand \citenamefont [1]{#1}%
\providecommand \href@noop [0]{\@secondoftwo}%
\providecommand \href [0]{\begingroup \@sanitize@url \@href}%
\providecommand \@href[1]{\@@startlink{#1}\@@href}%
\providecommand \@@href[1]{\endgroup#1\@@endlink}%
\providecommand \@sanitize@url [0]{\catcode `\\12\catcode `\$12\catcode
  `\&12\catcode `\#12\catcode `\^12\catcode `\_12\catcode `\%12\relax}%
\providecommand \@@startlink[1]{}%
\providecommand \@@endlink[0]{}%
\providecommand \url  [0]{\begingroup\@sanitize@url \@url }%
\providecommand \@url [1]{\endgroup\@href {#1}{\urlprefix }}%
\providecommand \urlprefix  [0]{URL }%
\providecommand \Eprint [0]{\href }%
\providecommand \doibase [0]{http://dx.doi.org/}%
\providecommand \selectlanguage [0]{\@gobble}%
\providecommand \bibinfo  [0]{\@secondoftwo}%
\providecommand \bibfield  [0]{\@secondoftwo}%
\providecommand \translation [1]{[#1]}%
\providecommand \BibitemOpen [0]{}%
\providecommand \bibitemStop [0]{}%
\providecommand \bibitemNoStop [0]{.\EOS\space}%
\providecommand \EOS [0]{\spacefactor3000\relax}%
\providecommand \BibitemShut  [1]{\csname bibitem#1\endcsname}%
\let\auto@bib@innerbib\@empty
%</preamble>
\bibitem [{\citenamefont {Aad}\ \emph {et~al.}(2012)\citenamefont {Aad},
  \citenamefont {Abajyan}, \citenamefont {Abbott}, \citenamefont {Abdallah},
  \citenamefont {Khalek}, \citenamefont {Abdelalim}, \citenamefont {Abdinov},
  \citenamefont {Aben}, \citenamefont {Abi}, \citenamefont {Abolins} \emph
  {et~al.}}]{Aad20121}%
  \BibitemOpen
  \bibfield  {author} {\bibinfo {author} {\bibfnamefont {G.}~\bibnamefont
  {Aad}}, \bibinfo {author} {\bibfnamefont {T.}~\bibnamefont {Abajyan}},
  \bibinfo {author} {\bibfnamefont {B.}~\bibnamefont {Abbott}}, \bibinfo
  {author} {\bibfnamefont {J.}~\bibnamefont {Abdallah}}, \bibinfo {author}
  {\bibfnamefont {S.~A.}\ \bibnamefont {Khalek}}, \bibinfo {author}
  {\bibfnamefont {A.}~\bibnamefont {Abdelalim}}, \bibinfo {author}
  {\bibfnamefont {O.}~\bibnamefont {Abdinov}}, \bibinfo {author} {\bibfnamefont
  {R.}~\bibnamefont {Aben}}, \bibinfo {author} {\bibfnamefont {B.}~\bibnamefont
  {Abi}}, \bibinfo {author} {\bibfnamefont {M.}~\bibnamefont {Abolins}},  \emph
  {et~al.},\ }\href {\doibase 10.1016/j.physletb.2012.08.020} {\bibfield
  {journal} {\bibinfo  {journal} {Physics Letters B}\ }\textbf {\bibinfo
  {volume} {716}},\ \bibinfo {pages} {1 } (\bibinfo {year} {2012})}\BibitemShut
  {NoStop}%
\bibitem [{\citenamefont {Chatrchyan}\ \emph {et~al.}(2012)\citenamefont
  {Chatrchyan}, \citenamefont {Khachatryan}, \citenamefont {Sirunyan},
  \citenamefont {Tumasyan}, \citenamefont {Adam}, \citenamefont {Aguilo},
  \citenamefont {Bergauer}, \citenamefont {Dragicevic}, \citenamefont {Erö},
  \citenamefont {Fabjan} \emph {et~al.}}]{Chatrchyan201230}%
  \BibitemOpen
  \bibfield  {author} {\bibinfo {author} {\bibfnamefont {S.}~\bibnamefont
  {Chatrchyan}}, \bibinfo {author} {\bibfnamefont {V.}~\bibnamefont
  {Khachatryan}}, \bibinfo {author} {\bibfnamefont {A.}~\bibnamefont
  {Sirunyan}}, \bibinfo {author} {\bibfnamefont {A.}~\bibnamefont {Tumasyan}},
  \bibinfo {author} {\bibfnamefont {W.}~\bibnamefont {Adam}}, \bibinfo {author}
  {\bibfnamefont {E.}~\bibnamefont {Aguilo}}, \bibinfo {author} {\bibfnamefont
  {T.}~\bibnamefont {Bergauer}}, \bibinfo {author} {\bibfnamefont
  {M.}~\bibnamefont {Dragicevic}}, \bibinfo {author} {\bibfnamefont
  {J.}~\bibnamefont {Erö}}, \bibinfo {author} {\bibfnamefont {C.}~\bibnamefont
  {Fabjan}},  \emph {et~al.},\ }\href {\doibase 10.1016/j.physletb.2012.08.021}
  {\bibfield  {journal} {\bibinfo  {journal} {Physics Letters B}\ }\textbf
  {\bibinfo {volume} {716}},\ \bibinfo {pages} {30 } (\bibinfo {year}
  {2012})}\BibitemShut {NoStop}%
\bibitem [{\citenamefont {Sher}(1989)}]{SHER1989273}%
  \BibitemOpen
  \bibfield  {author} {\bibinfo {author} {\bibfnamefont {M.}~\bibnamefont
  {Sher}},\ }\href {\doibase 10.1016/0370-1573(89)90061-6} {\bibfield
  {journal} {\bibinfo  {journal} {Physics Reports}\ }\textbf {\bibinfo {volume}
  {179}},\ \bibinfo {pages} {273 } (\bibinfo {year} {1989})}\BibitemShut
  {NoStop}%
\bibitem [{\citenamefont {Olive}\ \emph {et~al.}(2014)\citenamefont {Olive}
  \emph {et~al.}}]{Agashe:2014kda}%
  \BibitemOpen
  \bibfield  {author} {\bibinfo {author} {\bibfnamefont {K.~A.}\ \bibnamefont
  {Olive}} \emph {et~al.} (\bibinfo {collaboration} {Particle Data Group}),\
  }\href {\doibase 10.1088/1674-1137/38/9/090001} {\bibfield  {journal}
  {\bibinfo  {journal} {Chin. Phys.}\ }\textbf {\bibinfo {volume} {C38}},\
  \bibinfo {pages} {090001} (\bibinfo {year} {2014})}\BibitemShut {NoStop}%
%%CITATION = CHPHD,C38,090001;%%
\bibitem [{\citenamefont {Buttazzo}\ \emph {et~al.}(2013)\citenamefont
  {Buttazzo}, \citenamefont {Degrassi}, \citenamefont {Giardino}, \citenamefont
  {Giudice}, \citenamefont {Sala}, \citenamefont {Salvio},\ and\ \citenamefont
  {Strumia}}]{Buttazzo2013}%
  \BibitemOpen
  \bibfield  {author} {\bibinfo {author} {\bibfnamefont {D.}~\bibnamefont
  {Buttazzo}}, \bibinfo {author} {\bibfnamefont {G.}~\bibnamefont {Degrassi}},
  \bibinfo {author} {\bibfnamefont {P.~P.}\ \bibnamefont {Giardino}}, \bibinfo
  {author} {\bibfnamefont {G.~F.}\ \bibnamefont {Giudice}}, \bibinfo {author}
  {\bibfnamefont {F.}~\bibnamefont {Sala}}, \bibinfo {author} {\bibfnamefont
  {A.}~\bibnamefont {Salvio}}, \ and\ \bibinfo {author} {\bibfnamefont
  {A.}~\bibnamefont {Strumia}},\ }\href {\doibase 10.1007/JHEP12(2013)089}
  {\bibfield  {journal} {\bibinfo  {journal} {Journal of High Energy Physics}\
  }\textbf {\bibinfo {volume} {2013}},\ \bibinfo {pages} {1} (\bibinfo {year}
  {2013})}\BibitemShut {NoStop}%
\bibitem [{\citenamefont {Alekhin}\ \emph {et~al.}(2012)\citenamefont
  {Alekhin}, \citenamefont {Djouadi},\ and\ \citenamefont
  {Moch}}]{Alekhin2012214}%
  \BibitemOpen
  \bibfield  {author} {\bibinfo {author} {\bibfnamefont {S.}~\bibnamefont
  {Alekhin}}, \bibinfo {author} {\bibfnamefont {A.}~\bibnamefont {Djouadi}}, \
  and\ \bibinfo {author} {\bibfnamefont {S.}~\bibnamefont {Moch}},\ }\href
  {\doibase 10.1016/j.physletb.2012.08.024} {\bibfield  {journal} {\bibinfo
  {journal} {Physics Letters B}\ }\textbf {\bibinfo {volume} {716}},\ \bibinfo
  {pages} {214 } (\bibinfo {year} {2012})}\BibitemShut {NoStop}%
\bibitem [{\citenamefont {Degrassi}\ \emph {et~al.}(2012)\citenamefont
  {Degrassi}, \citenamefont {Di~Vita}, \citenamefont {Elias-Mir{\'o}},
  \citenamefont {Espinosa}, \citenamefont {Giudice}, \citenamefont {Isidori},\
  and\ \citenamefont {Strumia}}]{Degrassi2012}%
  \BibitemOpen
  \bibfield  {author} {\bibinfo {author} {\bibfnamefont {G.}~\bibnamefont
  {Degrassi}}, \bibinfo {author} {\bibfnamefont {S.}~\bibnamefont {Di~Vita}},
  \bibinfo {author} {\bibfnamefont {J.}~\bibnamefont {Elias-Mir{\'o}}},
  \bibinfo {author} {\bibfnamefont {J.~R.}\ \bibnamefont {Espinosa}}, \bibinfo
  {author} {\bibfnamefont {G.~F.}\ \bibnamefont {Giudice}}, \bibinfo {author}
  {\bibfnamefont {G.}~\bibnamefont {Isidori}}, \ and\ \bibinfo {author}
  {\bibfnamefont {A.}~\bibnamefont {Strumia}},\ }\href {\doibase
  10.1007/JHEP08(2012)098} {\bibfield  {journal} {\bibinfo  {journal} {Journal
  of High Energy Physics}\ }\textbf {\bibinfo {volume} {2012}},\ \bibinfo
  {pages} {1} (\bibinfo {year} {2012})}\BibitemShut {NoStop}%
\bibitem [{\citenamefont {Isidori}\ \emph {et~al.}(2008)\citenamefont
  {Isidori}, \citenamefont {Rychkov}, \citenamefont {Strumia},\ and\
  \citenamefont {Tetradis}}]{PhysRevD.77.025034}%
  \BibitemOpen
  \bibfield  {author} {\bibinfo {author} {\bibfnamefont {G.}~\bibnamefont
  {Isidori}}, \bibinfo {author} {\bibfnamefont {V.~S.}\ \bibnamefont
  {Rychkov}}, \bibinfo {author} {\bibfnamefont {A.}~\bibnamefont {Strumia}}, \
  and\ \bibinfo {author} {\bibfnamefont {N.}~\bibnamefont {Tetradis}},\ }\href
  {\doibase 10.1103/PhysRevD.77.025034} {\bibfield  {journal} {\bibinfo
  {journal} {Phys. Rev. D}\ }\textbf {\bibinfo {volume} {77}},\ \bibinfo
  {pages} {025034} (\bibinfo {year} {2008})}\BibitemShut {NoStop}%
\bibitem [{\citenamefont {Rajantie}\ and\ \citenamefont
  {Stopyra}(2017)}]{PhysRevD.95.025008}%
  \BibitemOpen
  \bibfield  {author} {\bibinfo {author} {\bibfnamefont {A.}~\bibnamefont
  {Rajantie}}\ and\ \bibinfo {author} {\bibfnamefont {S.}~\bibnamefont
  {Stopyra}},\ }\href {\doibase 10.1103/PhysRevD.95.025008} {\bibfield
  {journal} {\bibinfo  {journal} {Phys. Rev. D}\ }\textbf {\bibinfo {volume}
  {95}},\ \bibinfo {pages} {025008} (\bibinfo {year} {2017})}\BibitemShut
  {NoStop}%
\bibitem [{\citenamefont {Salvio}\ \emph {et~al.}(2016)\citenamefont {Salvio},
  \citenamefont {Strumia}, \citenamefont {Tetradis},\ and\ \citenamefont
  {Urbano}}]{Salvio2016}%
  \BibitemOpen
  \bibfield  {author} {\bibinfo {author} {\bibfnamefont {A.}~\bibnamefont
  {Salvio}}, \bibinfo {author} {\bibfnamefont {A.}~\bibnamefont {Strumia}},
  \bibinfo {author} {\bibfnamefont {N.}~\bibnamefont {Tetradis}}, \ and\
  \bibinfo {author} {\bibfnamefont {A.}~\bibnamefont {Urbano}},\ }\href
  {\doibase 10.1007/JHEP09(2016)054} {\bibfield  {journal} {\bibinfo  {journal}
  {Journal of High Energy Physics}\ }\textbf {\bibinfo {volume} {2016}},\
  \bibinfo {pages} {54} (\bibinfo {year} {2016})}\BibitemShut {NoStop}%
\bibitem [{\citenamefont {Kobakhidze}\ and\ \citenamefont
  {Spencer-Smith}(2013)}]{Kobakhidze2013130}%
  \BibitemOpen
  \bibfield  {author} {\bibinfo {author} {\bibfnamefont {A.}~\bibnamefont
  {Kobakhidze}}\ and\ \bibinfo {author} {\bibfnamefont {A.}~\bibnamefont
  {Spencer-Smith}},\ }\href {\doibase 10.1016/j.physletb.2013.04.013}
  {\bibfield  {journal} {\bibinfo  {journal} {Physics Letters B}\ }\textbf
  {\bibinfo {volume} {722}},\ \bibinfo {pages} {130 } (\bibinfo {year}
  {2013})}\BibitemShut {NoStop}%
\bibitem [{\citenamefont {Branchina}\ \emph {et~al.}(2014)\citenamefont
  {Branchina}, \citenamefont {Messina},\ and\ \citenamefont
  {Platania}}]{Branchina2014}%
  \BibitemOpen
  \bibfield  {author} {\bibinfo {author} {\bibfnamefont {V.}~\bibnamefont
  {Branchina}}, \bibinfo {author} {\bibfnamefont {E.}~\bibnamefont {Messina}},
  \ and\ \bibinfo {author} {\bibfnamefont {A.}~\bibnamefont {Platania}},\
  }\href {\doibase 10.1007/JHEP09(2014)182} {\bibfield  {journal} {\bibinfo
  {journal} {Journal of High Energy Physics}\ }\textbf {\bibinfo {volume}
  {2014}},\ \bibinfo {pages} {1} (\bibinfo {year} {2014})}\BibitemShut
  {NoStop}%
\bibitem [{\citenamefont {Fairbairn}\ and\ \citenamefont
  {Hogan}(2014)}]{PhysRevLett.112.201801}%
  \BibitemOpen
  \bibfield  {author} {\bibinfo {author} {\bibfnamefont {M.}~\bibnamefont
  {Fairbairn}}\ and\ \bibinfo {author} {\bibfnamefont {R.}~\bibnamefont
  {Hogan}},\ }\href {\doibase 10.1103/PhysRevLett.112.201801} {\bibfield
  {journal} {\bibinfo  {journal} {Phys. Rev. Lett.}\ }\textbf {\bibinfo
  {volume} {112}},\ \bibinfo {pages} {201801} (\bibinfo {year}
  {2014})}\BibitemShut {NoStop}%
\bibitem [{\citenamefont {Herranen}\ \emph {et~al.}(2014)\citenamefont
  {Herranen}, \citenamefont {Markkanen}, \citenamefont {Nurmi},\ and\
  \citenamefont {Rajantie}}]{PhysRevLett.113.211102}%
  \BibitemOpen
  \bibfield  {author} {\bibinfo {author} {\bibfnamefont {M.}~\bibnamefont
  {Herranen}}, \bibinfo {author} {\bibfnamefont {T.}~\bibnamefont {Markkanen}},
  \bibinfo {author} {\bibfnamefont {S.}~\bibnamefont {Nurmi}}, \ and\ \bibinfo
  {author} {\bibfnamefont {A.}~\bibnamefont {Rajantie}},\ }\href {\doibase
  10.1103/PhysRevLett.113.211102} {\bibfield  {journal} {\bibinfo  {journal}
  {Phys. Rev. Lett.}\ }\textbf {\bibinfo {volume} {113}},\ \bibinfo {pages}
  {211102} (\bibinfo {year} {2014})}\BibitemShut {NoStop}%
\bibitem [{\citenamefont {Herranen}\ \emph {et~al.}(2015)\citenamefont
  {Herranen}, \citenamefont {Markkanen}, \citenamefont {Nurmi},\ and\
  \citenamefont {Rajantie}}]{PhysRevLett.115.241301}%
  \BibitemOpen
  \bibfield  {author} {\bibinfo {author} {\bibfnamefont {M.}~\bibnamefont
  {Herranen}}, \bibinfo {author} {\bibfnamefont {T.}~\bibnamefont {Markkanen}},
  \bibinfo {author} {\bibfnamefont {S.}~\bibnamefont {Nurmi}}, \ and\ \bibinfo
  {author} {\bibfnamefont {A.}~\bibnamefont {Rajantie}},\ }\href {\doibase
  10.1103/PhysRevLett.115.241301} {\bibfield  {journal} {\bibinfo  {journal}
  {Phys. Rev. Lett.}\ }\textbf {\bibinfo {volume} {115}},\ \bibinfo {pages}
  {241301} (\bibinfo {year} {2015})}\BibitemShut {NoStop}%
\bibitem [{\citenamefont {Calmet}\ \emph {et~al.}(2017)\citenamefont {Calmet},
  \citenamefont {Kuntz},\ and\ \citenamefont {Moss}}]{Calmet:2017hja}%
  \BibitemOpen
  \bibfield  {author} {\bibinfo {author} {\bibfnamefont {X.}~\bibnamefont
  {Calmet}}, \bibinfo {author} {\bibfnamefont {I.}~\bibnamefont {Kuntz}}, \
  and\ \bibinfo {author} {\bibfnamefont {I.~G.}\ \bibnamefont {Moss}},\
  }\href@noop {} {\  (\bibinfo {year} {2017})},\ \Eprint
  {http://arxiv.org/abs/1701.02140} {arXiv:1701.02140 [hep-ph]} \BibitemShut
  {NoStop}%
%%CITATION = ARXIV:1701.02140;%%
\bibitem [{\citenamefont {Markkanen}\ \emph {et~al.}(2017)\citenamefont
  {Markkanen}, \citenamefont {Nurmi},\ and\ \citenamefont
  {Rajantie}}]{Markkanen:2017dlc}%
  \BibitemOpen
  \bibfield  {author} {\bibinfo {author} {\bibfnamefont {T.}~\bibnamefont
  {Markkanen}}, \bibinfo {author} {\bibfnamefont {S.}~\bibnamefont {Nurmi}}, \
  and\ \bibinfo {author} {\bibfnamefont {A.}~\bibnamefont {Rajantie}},\
  }\href@noop {} {\  (\bibinfo {year} {2017})},\ \Eprint
  {http://arxiv.org/abs/1707.00866} {arXiv:1707.00866 [hep-ph]} \BibitemShut
  {NoStop}%
%%CITATION = ARXIV:1707.00866;%%
\bibitem [{\citenamefont {Espinosa}\ \emph {et~al.}(2015)\citenamefont
  {Espinosa}, \citenamefont {Giudice}, \citenamefont {Morgante}, \citenamefont
  {Riotto}, \citenamefont {Senatore}, \citenamefont {Strumia},\ and\
  \citenamefont {Tetradis}}]{Espinosa2015}%
  \BibitemOpen
  \bibfield  {author} {\bibinfo {author} {\bibfnamefont {J.~R.}\ \bibnamefont
  {Espinosa}}, \bibinfo {author} {\bibfnamefont {G.~F.}\ \bibnamefont
  {Giudice}}, \bibinfo {author} {\bibfnamefont {E.}~\bibnamefont {Morgante}},
  \bibinfo {author} {\bibfnamefont {A.}~\bibnamefont {Riotto}}, \bibinfo
  {author} {\bibfnamefont {L.}~\bibnamefont {Senatore}}, \bibinfo {author}
  {\bibfnamefont {A.}~\bibnamefont {Strumia}}, \ and\ \bibinfo {author}
  {\bibfnamefont {N.}~\bibnamefont {Tetradis}},\ }\href {\doibase
  10.1007/JHEP09(2015)174} {\bibfield  {journal} {\bibinfo  {journal} {Journal
  of High Energy Physics}\ }\textbf {\bibinfo {volume} {2015}},\ \bibinfo
  {pages} {174} (\bibinfo {year} {2015})}\BibitemShut {NoStop}%
\bibitem [{\citenamefont {Coleman}\ and\ \citenamefont
  {De~Luccia}(1980)}]{PhysRevD.21.3305}%
  \BibitemOpen
  \bibfield  {author} {\bibinfo {author} {\bibfnamefont {S.}~\bibnamefont
  {Coleman}}\ and\ \bibinfo {author} {\bibfnamefont {F.}~\bibnamefont
  {De~Luccia}},\ }\href {\doibase 10.1103/PhysRevD.21.3305} {\bibfield
  {journal} {\bibinfo  {journal} {Phys. Rev. D}\ }\textbf {\bibinfo {volume}
  {21}},\ \bibinfo {pages} {3305} (\bibinfo {year} {1980})}\BibitemShut
  {NoStop}%
\bibitem [{\citenamefont {Joti}\ \emph {et~al.}(2017)\citenamefont {Joti},
  \citenamefont {Katsis}, \citenamefont {Loupas}, \citenamefont {Salvio},
  \citenamefont {Strumia}, \citenamefont {Tetradis},\ and\ \citenamefont
  {Urbano}}]{Joti2017}%
  \BibitemOpen
  \bibfield  {author} {\bibinfo {author} {\bibfnamefont {A.}~\bibnamefont
  {Joti}}, \bibinfo {author} {\bibfnamefont {A.}~\bibnamefont {Katsis}},
  \bibinfo {author} {\bibfnamefont {D.}~\bibnamefont {Loupas}}, \bibinfo
  {author} {\bibfnamefont {A.}~\bibnamefont {Salvio}}, \bibinfo {author}
  {\bibfnamefont {A.}~\bibnamefont {Strumia}}, \bibinfo {author} {\bibfnamefont
  {N.}~\bibnamefont {Tetradis}}, \ and\ \bibinfo {author} {\bibfnamefont
  {A.}~\bibnamefont {Urbano}},\ }\href {\doibase 10.1007/JHEP07(2017)058}
  {\bibfield  {journal} {\bibinfo  {journal} {Journal of High Energy Physics}\
  }\textbf {\bibinfo {volume} {2017}},\ \bibinfo {pages} {58} (\bibinfo {year}
  {2017})}\BibitemShut {NoStop}%
\bibitem [{\citenamefont {Starobinsky}(1986)}]{Starobinsky1986}%
  \BibitemOpen
  \bibfield  {author} {\bibinfo {author} {\bibfnamefont {A.~A.}\ \bibnamefont
  {Starobinsky}},\ }\enquote {\bibinfo {title} {Stochastic de sitter
  (inflationary) stage in the early universe},}\ in\ \href {\doibase
  10.1007/3-540-16452-9_6} {\emph {\bibinfo {booktitle} {Field Theory, Quantum
  Gravity and Strings: Proceedings of a Seminar Series Held at DAPHE,
  Observatoire de Meudon, and LPTHE, Universit{\'e} Pierre et Marie Curie,
  Paris, Between October 1984 and October 1985}}},\ \bibinfo {editor} {edited
  by\ \bibinfo {editor} {\bibfnamefont {H.~J.}\ \bibnamefont {de~Vega}}\ and\
  \bibinfo {editor} {\bibfnamefont {N.}~\bibnamefont {S{\'a}nchez}}}\ (\bibinfo
   {publisher} {Springer Berlin Heidelberg},\ \bibinfo {address} {Berlin,
  Heidelberg},\ \bibinfo {year} {1986})\ pp.\ \bibinfo {pages}
  {107--126}\BibitemShut {NoStop}%
\bibitem [{\citenamefont {Hawking}\ and\ \citenamefont
  {Moss}(1982)}]{HAWKING198235}%
  \BibitemOpen
  \bibfield  {author} {\bibinfo {author} {\bibfnamefont {S.}~\bibnamefont
  {Hawking}}\ and\ \bibinfo {author} {\bibfnamefont {I.}~\bibnamefont {Moss}},\
  }\href {\doibase http://dx.doi.org/10.1016/0370-2693(82)90946-7} {\bibfield
  {journal} {\bibinfo  {journal} {Physics Letters B}\ }\textbf {\bibinfo
  {volume} {110}},\ \bibinfo {pages} {35 } (\bibinfo {year}
  {1982})}\BibitemShut {NoStop}%
\bibitem [{\citenamefont {Coleman}(1977)}]{PhysRevD.15.2929}%
  \BibitemOpen
  \bibfield  {author} {\bibinfo {author} {\bibfnamefont {S.}~\bibnamefont
  {Coleman}},\ }\href {\doibase 10.1103/PhysRevD.15.2929} {\bibfield  {journal}
  {\bibinfo  {journal} {Phys. Rev. D}\ }\textbf {\bibinfo {volume} {15}},\
  \bibinfo {pages} {2929} (\bibinfo {year} {1977})}\BibitemShut {NoStop}%
\bibitem [{\citenamefont {Callan}\ and\ \citenamefont
  {Coleman}(1977)}]{PhysRevD.16.1762}%
  \BibitemOpen
  \bibfield  {author} {\bibinfo {author} {\bibfnamefont {C.~G.}\ \bibnamefont
  {Callan}}\ and\ \bibinfo {author} {\bibfnamefont {S.}~\bibnamefont
  {Coleman}},\ }\href {\doibase 10.1103/PhysRevD.16.1762} {\bibfield  {journal}
  {\bibinfo  {journal} {Phys. Rev. D}\ }\textbf {\bibinfo {volume} {16}},\
  \bibinfo {pages} {1762} (\bibinfo {year} {1977})}\BibitemShut {NoStop}%
\bibitem [{\citenamefont {Brown}\ and\ \citenamefont
  {Weinberg}(2007)}]{PhysRevD.76.064003}%
  \BibitemOpen
  \bibfield  {author} {\bibinfo {author} {\bibfnamefont {A.~R.}\ \bibnamefont
  {Brown}}\ and\ \bibinfo {author} {\bibfnamefont {E.~J.}\ \bibnamefont
  {Weinberg}},\ }\href {\doibase 10.1103/PhysRevD.76.064003} {\bibfield
  {journal} {\bibinfo  {journal} {Phys. Rev. D}\ }\textbf {\bibinfo {volume}
  {76}},\ \bibinfo {pages} {064003} (\bibinfo {year} {2007})}\BibitemShut
  {NoStop}%
\bibitem [{\citenamefont {Gibbons}\ and\ \citenamefont
  {Hawking}(1977)}]{PhysRevD.15.2752}%
  \BibitemOpen
  \bibfield  {author} {\bibinfo {author} {\bibfnamefont {G.~W.}\ \bibnamefont
  {Gibbons}}\ and\ \bibinfo {author} {\bibfnamefont {S.~W.}\ \bibnamefont
  {Hawking}},\ }\href {\doibase 10.1103/PhysRevD.15.2752} {\bibfield  {journal}
  {\bibinfo  {journal} {Phys. Rev. D}\ }\textbf {\bibinfo {volume} {15}},\
  \bibinfo {pages} {2752} (\bibinfo {year} {1977})}\BibitemShut {NoStop}%
\bibitem [{\citenamefont {Coleman}(1988)}]{COLEMAN1988178}%
  \BibitemOpen
  \bibfield  {author} {\bibinfo {author} {\bibfnamefont {S.}~\bibnamefont
  {Coleman}},\ }\href {\doibase http://dx.doi.org/10.1016/0550-3213(88)90308-2}
  {\bibfield  {journal} {\bibinfo  {journal} {Nuclear Physics B}\ }\textbf
  {\bibinfo {volume} {298}},\ \bibinfo {pages} {178 } (\bibinfo {year}
  {1988})}\BibitemShut {NoStop}%
\bibitem [{\citenamefont {Gratton}\ and\ \citenamefont
  {Turok}(2001)}]{PhysRevD.63.123514}%
  \BibitemOpen
  \bibfield  {author} {\bibinfo {author} {\bibfnamefont {S.}~\bibnamefont
  {Gratton}}\ and\ \bibinfo {author} {\bibfnamefont {N.}~\bibnamefont
  {Turok}},\ }\href {\doibase 10.1103/PhysRevD.63.123514} {\bibfield  {journal}
  {\bibinfo  {journal} {Phys. Rev. D}\ }\textbf {\bibinfo {volume} {63}},\
  \bibinfo {pages} {123514} (\bibinfo {year} {2001})}\BibitemShut {NoStop}%
\bibitem [{\citenamefont {Balek}\ and\ \citenamefont
  {Demetrian}(2004)}]{PhysRevD.69.063518}%
  \BibitemOpen
  \bibfield  {author} {\bibinfo {author} {\bibfnamefont {V.}~\bibnamefont
  {Balek}}\ and\ \bibinfo {author} {\bibfnamefont {M.}~\bibnamefont
  {Demetrian}},\ }\href {\doibase 10.1103/PhysRevD.69.063518} {\bibfield
  {journal} {\bibinfo  {journal} {Phys. Rev. D}\ }\textbf {\bibinfo {volume}
  {69}},\ \bibinfo {pages} {063518} (\bibinfo {year} {2004})}\BibitemShut
  {NoStop}%
\bibitem [{\citenamefont {Coleman}\ \emph {et~al.}(1978)\citenamefont
  {Coleman}, \citenamefont {Glaser},\ and\ \citenamefont
  {Martin}}]{ColemanGlaserMartin1978}%
  \BibitemOpen
  \bibfield  {author} {\bibinfo {author} {\bibfnamefont {S.}~\bibnamefont
  {Coleman}}, \bibinfo {author} {\bibfnamefont {V.}~\bibnamefont {Glaser}}, \
  and\ \bibinfo {author} {\bibfnamefont {A.}~\bibnamefont {Martin}},\ }\href
  {\doibase 10.1007/BF01609421} {\bibfield  {journal} {\bibinfo  {journal}
  {Communications in Mathematical Physics}\ }\textbf {\bibinfo {volume} {58}},\
  \bibinfo {pages} {211} (\bibinfo {year} {1978})}\BibitemShut {NoStop}%
\bibitem [{\citenamefont {Garriga}\ and\ \citenamefont
  {Megevand}(2004)}]{Garriga:2004nm}%
  \BibitemOpen
  \bibfield  {author} {\bibinfo {author} {\bibfnamefont {J.}~\bibnamefont
  {Garriga}}\ and\ \bibinfo {author} {\bibfnamefont {A.}~\bibnamefont
  {Megevand}},\ }\bibfield  {booktitle} {\emph {\bibinfo {booktitle} {{The
  early universe: Confronting theory with observations. Proceedings, 8th
  Workshop, Peyresq Physics 8, Peyresq, France, June 21-27, 2003}}},\ }\href
  {\doibase 10.1023/B:IJTP.0000048178.69097.fb} {\bibfield  {journal} {\bibinfo
   {journal} {Int. J. Theor. Phys.}\ }\textbf {\bibinfo {volume} {43}},\
  \bibinfo {pages} {883} (\bibinfo {year} {2004})},\ \Eprint
  {http://arxiv.org/abs/hep-th/0404097} {arXiv:hep-th/0404097 [hep-th]}
  \BibitemShut {NoStop}%
%%CITATION = HEP-TH/0404097;%%
\bibitem [{\citenamefont {Masoumi}\ and\ \citenamefont
  {Weinberg}(2012)}]{Masoumi:2012yy}%
  \BibitemOpen
  \bibfield  {author} {\bibinfo {author} {\bibfnamefont {A.}~\bibnamefont
  {Masoumi}}\ and\ \bibinfo {author} {\bibfnamefont {E.~J.}\ \bibnamefont
  {Weinberg}},\ }\href {\doibase 10.1103/PhysRevD.86.104029} {\bibfield
  {journal} {\bibinfo  {journal} {Phys. Rev.}\ }\textbf {\bibinfo {volume}
  {D86}},\ \bibinfo {pages} {104029} (\bibinfo {year} {2012})},\ \Eprint
  {http://arxiv.org/abs/1207.3717} {arXiv:1207.3717 [hep-th]} \BibitemShut
  {NoStop}%
%%CITATION = ARXIV:1207.3717;%%
\bibitem [{\citenamefont {Hackworth}\ and\ \citenamefont
  {Weinberg}(2005)}]{PhysRevD.71.044014}%
  \BibitemOpen
  \bibfield  {author} {\bibinfo {author} {\bibfnamefont {J.~C.}\ \bibnamefont
  {Hackworth}}\ and\ \bibinfo {author} {\bibfnamefont {E.~J.}\ \bibnamefont
  {Weinberg}},\ }\href {\doibase 10.1103/PhysRevD.71.044014} {\bibfield
  {journal} {\bibinfo  {journal} {Phys. Rev. D}\ }\textbf {\bibinfo {volume}
  {71}},\ \bibinfo {pages} {044014} (\bibinfo {year} {2005})}\BibitemShut
  {NoStop}%
\bibitem [{\citenamefont {Lavrelashvili}(2006)}]{PhysRevD.73.083513}%
  \BibitemOpen
  \bibfield  {author} {\bibinfo {author} {\bibfnamefont {G.}~\bibnamefont
  {Lavrelashvili}},\ }\href {\doibase 10.1103/PhysRevD.73.083513} {\bibfield
  {journal} {\bibinfo  {journal} {Phys. Rev. D}\ }\textbf {\bibinfo {volume}
  {73}},\ \bibinfo {pages} {083513} (\bibinfo {year} {2006})}\BibitemShut
  {NoStop}%
\bibitem [{\citenamefont {Battarra}\ \emph {et~al.}(2012)\citenamefont
  {Battarra}, \citenamefont {Lavrelashvili},\ and\ \citenamefont
  {Lehners}}]{PhysRevD.86.124001}%
  \BibitemOpen
  \bibfield  {author} {\bibinfo {author} {\bibfnamefont {L.}~\bibnamefont
  {Battarra}}, \bibinfo {author} {\bibfnamefont {G.}~\bibnamefont
  {Lavrelashvili}}, \ and\ \bibinfo {author} {\bibfnamefont {J.-L.}\
  \bibnamefont {Lehners}},\ }\href {\doibase 10.1103/PhysRevD.86.124001}
  {\bibfield  {journal} {\bibinfo  {journal} {Phys. Rev. D}\ }\textbf {\bibinfo
  {volume} {86}},\ \bibinfo {pages} {124001} (\bibinfo {year}
  {2012})}\BibitemShut {NoStop}%
\bibitem [{\citenamefont {Khvedelidze}\ \emph {et~al.}(2000)\citenamefont
  {Khvedelidze}, \citenamefont {Lavrelashvili},\ and\ \citenamefont
  {Tanaka}}]{PhysRevD.62.083501}%
  \BibitemOpen
  \bibfield  {author} {\bibinfo {author} {\bibfnamefont {A.}~\bibnamefont
  {Khvedelidze}}, \bibinfo {author} {\bibfnamefont {G.}~\bibnamefont
  {Lavrelashvili}}, \ and\ \bibinfo {author} {\bibfnamefont {T.}~\bibnamefont
  {Tanaka}},\ }\href {\doibase 10.1103/PhysRevD.62.083501} {\bibfield
  {journal} {\bibinfo  {journal} {Phys. Rev. D}\ }\textbf {\bibinfo {volume}
  {62}},\ \bibinfo {pages} {083501} (\bibinfo {year} {2000})}\BibitemShut
  {NoStop}%
\bibitem [{\citenamefont {Jensen}\ and\ \citenamefont
  {Steinhardt}(1984)}]{Jensen1984176}%
  \BibitemOpen
  \bibfield  {author} {\bibinfo {author} {\bibfnamefont {L.~G.}\ \bibnamefont
  {Jensen}}\ and\ \bibinfo {author} {\bibfnamefont {P.~J.}\ \bibnamefont
  {Steinhardt}},\ }\href {\doibase 10.1016/0550-3213(84)90021-X} {\bibfield
  {journal} {\bibinfo  {journal} {Nuclear Physics B}\ }\textbf {\bibinfo
  {volume} {237}},\ \bibinfo {pages} {176 } (\bibinfo {year}
  {1984})}\BibitemShut {NoStop}%
\bibitem [{\citenamefont {Isidori}\ \emph {et~al.}(2001)\citenamefont
  {Isidori}, \citenamefont {Ridolfi},\ and\ \citenamefont
  {Strumia}}]{Isidori2001387}%
  \BibitemOpen
  \bibfield  {author} {\bibinfo {author} {\bibfnamefont {G.}~\bibnamefont
  {Isidori}}, \bibinfo {author} {\bibfnamefont {G.}~\bibnamefont {Ridolfi}}, \
  and\ \bibinfo {author} {\bibfnamefont {A.}~\bibnamefont {Strumia}},\ }\href
  {\doibase 10.1016/S0550-3213(01)00302-9} {\bibfield  {journal} {\bibinfo
  {journal} {Nuclear Physics B}\ }\textbf {\bibinfo {volume} {609}},\ \bibinfo
  {pages} {387 } (\bibinfo {year} {2001})}\BibitemShut {NoStop}%
\bibitem [{\citenamefont {Fehlberg}(1968)}]{Fehlberg78}%
  \BibitemOpen
  \bibfield  {author} {\bibinfo {author} {\bibfnamefont {E.}~\bibnamefont
  {Fehlberg}},\ }\href@noop {} {\emph {\bibinfo {title} {Classical Fifth-,
  Sixth-, Seventh-, and Eighth-Order Runge-Kutta Formulas with Stepsize
  Control}}},\ \bibinfo {type} {Technical Report}\ \bibinfo {number}
  {19680027281}\ (\bibinfo  {institution} {NASA Marshall Space Flight Center},\
  \bibinfo {address} {Huntsville, AL, United States},\ \bibinfo {year}
  {1968})\BibitemShut {NoStop}%
\bibitem [{\citenamefont {{Planck Collaboration}}\ \emph
  {et~al.}(2016)\citenamefont {{Planck Collaboration}}, \citenamefont {{Ade, P.
  A. R.}}, \citenamefont {{Aghanim, N.}}, \citenamefont {{Arnaud, M.}},
  \citenamefont {{Arroja, F.}}, \citenamefont {{Ashdown, M.}}, \citenamefont
  {{Aumont, J.}}, \citenamefont {{Baccigalupi, C.}},\ and\ \citenamefont
  {{others}}}]{planck}%
  \BibitemOpen
  \bibfield  {author} {\bibinfo {author} {\bibnamefont {{Planck
  Collaboration}}}, \bibinfo {author} {\bibnamefont {{Ade, P. A. R.}}},
  \bibinfo {author} {\bibnamefont {{Aghanim, N.}}}, \bibinfo {author}
  {\bibnamefont {{Arnaud, M.}}}, \bibinfo {author} {\bibnamefont {{Arroja,
  F.}}}, \bibinfo {author} {\bibnamefont {{Ashdown, M.}}}, \bibinfo {author}
  {\bibnamefont {{Aumont, J.}}}, \bibinfo {author} {\bibnamefont {{Baccigalupi,
  C.}}}, \ and\ \bibinfo {author} {\bibnamefont {{others}}},\ }\href {\doibase
  10.1051/0004-6361/201525898} {\bibfield  {journal} {\bibinfo  {journal}
  {A\&A}\ }\textbf {\bibinfo {volume} {594}},\ \bibinfo {pages} {A20} (\bibinfo
  {year} {2016})}\BibitemShut {NoStop}%
\bibitem [{\citenamefont {Nielsen}(1975)}]{NIELSEN1975173}%
  \BibitemOpen
  \bibfield  {author} {\bibinfo {author} {\bibfnamefont {N.}~\bibnamefont
  {Nielsen}},\ }\href {\doibase http://dx.doi.org/10.1016/0550-3213(75)90301-6}
  {\bibfield  {journal} {\bibinfo  {journal} {Nuclear Physics B}\ }\textbf
  {\bibinfo {volume} {101}},\ \bibinfo {pages} {173 } (\bibinfo {year}
  {1975})}\BibitemShut {NoStop}%
\end{thebibliography}%

\end{document}